\documentclass[english,twocolumn,aps,prl,superscriptaddress,showpacs, amsmath]{revtex4-2}
\usepackage[utf8]{inputenc}
\usepackage{amsmath}
\usepackage{amssymb}
\usepackage{graphicx}
\usepackage{color}
\usepackage{varwidth}
\usepackage[bookmarks=false]{hyperref}
\usepackage[usenames,dvipsnames]{xcolor}
\usepackage[super]{nth}
\usepackage{ulem,lipsum}
\usepackage{subcaption} 
\usepackage[export]{adjustbox} 
\usepackage{cleveref} 
\usepackage{tikz}
\usetikzlibrary{shapes}
\usepackage{soul}
\begin{document}

\title{Optical control of orbital magnetism in magic angle twisted bilayer graphene}

\author{Eylon Persky}
\email{perskye1@stanford.edu}
\affiliation{Geballe Laboratory for Advanced Materials, Stanford University, Stanford, CA 94305, USA}
\affiliation{Stanford Institute for Materials and Energy Sciences, SLAC National Accelerator Laboratory, 2575 Sand Hill Road, Menlo Park, CA 94025, USA}
\affiliation{Department of Applied Physics, Stanford University, Stanford, CA 94305, USA}
\author{Minhao He}
\affiliation{Department of Physics, University of Washington, Seattle, Washington, 98195, USA}
\author{Jiaqi Cai}
\affiliation{Department of Physics, University of Washington, Seattle, Washington, 98195, USA}
\author{Takashi Taniguchi}
\affiliation{Research Center for Materials Nanoarchitectonics, National Institute for Materials Science, 1-1 Namiki, Tsukuba 305-0044, Japan}
\author{Kenji Watanabe}
\affiliation{Research Center for Electronic and Optical Materials, National Institute for Materials Science, 1-1 Namiki, Tsukuba 305-0044, Japan}
\author{Xiaodong Xu}
\affiliation{Department of Physics, University of Washington, Seattle, Washington, 98195, USA}
\author{Aharon Kapitulnik}
\affiliation{Geballe Laboratory for Advanced Materials, Stanford University, Stanford, CA 94305, USA}
\affiliation{Stanford Institute for Materials and Energy Sciences, SLAC National Accelerator Laboratory, 2575 Sand Hill Road, Menlo Park, CA 94025, USA}
\affiliation{Department of Applied Physics, Stanford University, Stanford, CA 94305, USA}
\affiliation{Department of Physics, Stanford University, Stanford, CA 94305, USA}

\begin{abstract}
Flat bands in graphene-based moir\'e structures host a wide range of emerging strongly correlated and topological phenomena. Optically probing and controlling them can reveal important information such as symmetry and dynamics, but have so far been challenging due to the small energy gap compared to optical wavelengths. Here, we report near infrared optical control of orbital magnetism and associated anomalous Hall effects (AHE) in a magic angle twisted bilayer graphene (MATBG) on monolayer WSe$_2$ device. We show that the properties of the AHE, such as hysteresis and amplitude, can be controlled by light near integer moir\'e fillings, where spontaneous ferromagnetism exists. By modulating the light helicity, we observe periodic modulation of the transverse resistance in a wide range of fillings, indicating light induced orbital magnetization through a large inverse Faraday effect. At the transition between metallic and AHE regimes, we also reveal large and random switching of the Hall resistivity, which are attributed to optical control of percolating cluster of magnetic domains. Our results open the door to optical manipulation of correlation and topology in MATBG and related structures. 
\end{abstract}

\date{\today}

\maketitle
\section{Introduction}
Magic angle twisted bilayer graphene (MATBG) exhibits a rich phase diagram of topological and strongly correlated phases including correlated insulators with nearby superconductivity \cite{Cao2018a,Cao2018b}, orbital magnetism \cite{Sharpe2019,Serlin2020}, fractional Chern insulators \cite{Xie2021} and charge density waves \cite{Jiang2019}. This remarkable diversity of electronic phases originates form a periodic moir\'e superlattice with a large unit cell of $\sim10$ nm, which gives rise to flat energy bands \cite{Bistritzer2011},  that are separated from the dispersive bands by energy gaps of $\sim$50 meV. Achieving sufficiently flat bands \cite{Bistritzer2011} typically requires a narrow range of twist angles, within $\sim10\%$ of $\sim1.1^\circ$, a  characteristic that often prevented  the observation of all these striking phenomena in the same sample. This situation changes when a monolayer WSe$_2$ is placed on top of the MATBG, which is argued to stabilize the twist angle and reduce the resulting moir\'e disorder \cite{Polski2022}. Indeed, recent transport studies on MATBG/WSe$_2$ system \cite{He2024} yielded transport results consistent with \textit{all} previous measurements on MATBG with or without WSe$_2$.  

While the unique electronic band structure of MATBG is expected to significantly influence its optical properties, this research direction has been mostly dominated by theory (e.g. see \cite{Liu2020,Liu2024}) due to the relatively weak interaction of light with MATBG and the difficulty to avoid thermal effects, especially at low temperatures. This issue is acutely evident when we compare MATBG to twisted homo- and hetero-bilayers of transition metal dichalcogenides (TMDs) \cite{Wu2018}. TMDs typically exhibit a semiconducting gap of $\sim$1 eV, corresponding to a wavelength of $\sim$ 1.1 $\mu$m, which allows for direct optical probing of electronic compressibility and circular dichroism \cite{Cai2023,Zeng2023} through coupling to resonant excitations such as excitons \cite{Chernikov2014}. By contrast, in twisted graphene structures, there is no such sharp resonance, suggesting that the interaction between the light and flat-band electrons is weak. Furthermore, the small separation of the flat bands from the remote bands means that optical light excites electrons to/from the remote bands \cite{Tielrooij2013,Gierz2013}, thus changing the effective flat band filling, and given the weak electron-phonon coupling in graphene, light-induced heating is expected to strongly affect low-temperature highly-correlated and topological states \cite{Battista2022,Merino2024}.  
Despite exciting theoretical predictions \cite{Pershoguba2022,Yang2023}, studies of MATBG structures under illumination have been limited mainly to exploring phenomena that persist to elevated temperatures such as broken structural symmetries through the bulk photovoltaic effect \cite{Kumar2024}, where heating is not significant.

Here, we study the transport properties of a MATBG/WSe$_2$ structure with a nominal twist angle of $\theta=1.03^\circ$, under illumination with 1.55 $\mu$m wavelength near infra-red (NIR) light, focusing on the anomalous Hall states near fillings $\nu=1$ and $\nu=2$.  We show that illumination broadens the gross features of the gate-dependent longitudinal resistance ($R_{xx}$), finding that the heating is uniform across all doping. Comparing to previous reports on MATBG without WSe$_2$ \cite{Battista2022,Merino2024,Kumar2024}, we reveal a reduced heating which suggests that the WSe$_2$ layer provides a route for improved thermalization processes. Using unpolarized light, we find that for $\nu=1$ the AHE persists when the sample is illuminated with considerable power, exhibiting enhanced signal and somewhat reduced coercive field.  By contrast, the $\nu=2$ state is highly sensitive to light, losing the coercive field even at 100 times smaller power than for the $\nu=1$, despite a similar Curie temperature \cite{He2024}, an effect we attribute to the nature of band filling and the emergence of the magnetic state near $\nu=2$. Focusing on the more robust $\nu=1$ state, we show that circularly polarized light can be used to control the magnetization state. Time dynamics of the light-induced switching reveal that the Hall resistance is dominated by a backbone of domains which percolate between the contacts, clarifying the microscopic picture the filling-dependent transition between the correlated insulator and the metallic states. Our results open the door to further optical investigations of graphene-based structures, which could manipulate and probe the symmetry properties of the various strongly correlated states.

\section{Results and analysis}
\subsection{Transport properties under illumination}
\begin{figure*}[ht!]
\centering
\includegraphics[width=\textwidth]{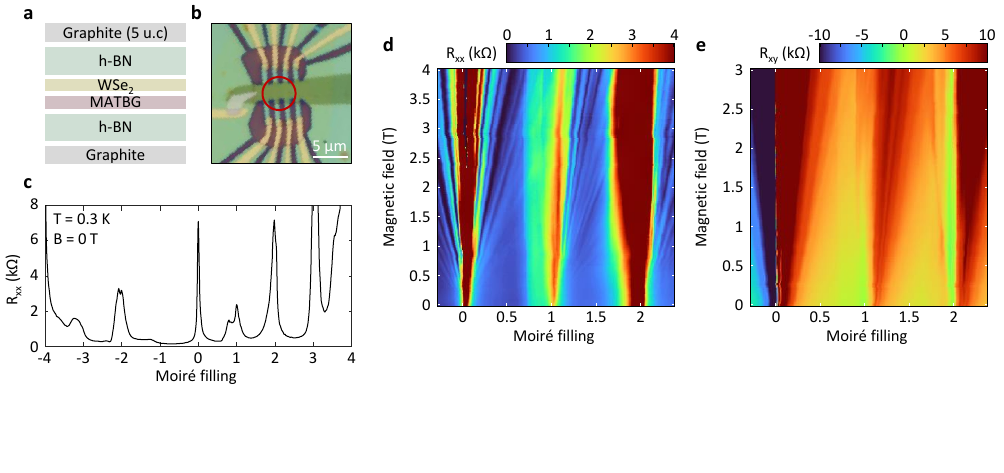}
\caption{Transport characterization of the sample. (a) Schematic of the device structure. (b) Optical microscope image of the device. The red circle is the $1/e^2$ spot size. (c) The longitudinal resistance, $R_{xx}$, as a function of filling, measured at zero field and at 300 mK, showing insulating peaks at the charge neutrality point and at integer fillings. A Resistive hump feature appears at fillings $0.7 \leq \nu \leq 1$. (d,e) The longitudinal (d) and transverse (e) resistance as a function of filling and magnetic field. The data show typical Landau fans that emerge from $\nu = 0$ and from the left side of $\nu = 1$ and $\nu = 2$. The hump feature at $\nu = 0.7$ does not disperse as a function of field, in both $R_{xx}$ and $R_{xy}.$}
\label{fig-photodoping}
\end{figure*}

We studied a MATBG/WSe$_2$ heterostructure (Fig. \ref{fig-photodoping}a,b) with top and bottom graphite gates. Figure \ref{fig-photodoping}c shows the longitudinal resistance, $R_{xx}$ as a function of filling factor measured at 0.3 K. Without illumination, the sample shows characteristic insulating peaks at integer fillings, along with superconductivity near $\nu = -2$ and an anomalous Hall effect near $\nu=1$ and $\nu=2$ \cite{He2024}. The AHE at an integer filling is due to spin-orbit coupling induced by the WSe$_2$ layer, which favors a valley-polarized state at half filling \cite{Lin2022}. Interestingly, there is a hump feature in $R_{xx}$ at fillings $0.7\leq \nu \leq 0.9$, before the value of $R_{xx}$ increases further and peaks at $\nu = 1$. To understand the origin of this hump feature, we investigated its dependence on the magnetic field. Figs. \ref{fig-photodoping}d,e show typical Landau fans that emanate from the integer fillings. The hump feature at $\nu = 0.7$ does not change with magnetic field, appearing as a vertical feature in both $R_{xx}$ and $R_{xy}$. This non-dispersing behavior may indicate an insulating state, in which one of the Fermi surfaces is gapped.

\begin{figure*}[ht!]
\centering
\includegraphics[width=\textwidth]{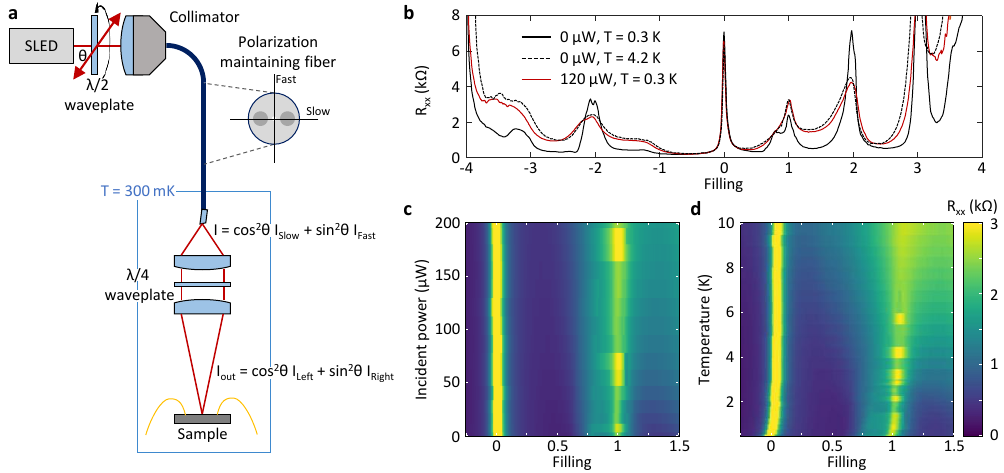}
\caption{Light-induced heating. (a) Schematic of the experimental setup. Unpolarized light (i.e. light of equal amount of circular polarization) is launched into the cryostat using a polarization maintaining fiber, and is then recollimated and focused on the sample at low temperatures. Rotating the half waveplate at room temperature changes the intensity ratio of left and right circularly polarized light incident on the sample.
(b) $R_{xx}$ as a function of filling, measured at 4 K and at 300 mK, with and without illuminating the sample. (c,d) $R_{xx}$ for the filling range $-0.25 \leq \nu \leq 1.5$ at increasing light intensities (c) and temperatures (d).}
\label{fig-heating}
\end{figure*}

Broadband NIR light (center wavelength- 1550 nm, 3dB bandwidth, 90 nm) was launched from a superluminescent Diode (SLED) at room temperature into a polarization maintaining fiber and focused onto a spot ($1/e^2$-diameter, 4.5 $\mu$m), such that the entire device area between the contacts was illuminated (Fig. \ref{fig-heating}a). Light traveling along the slow axis of the fiber acquires a delay with respect to the light traveling along the fast axis. Since the length of the fiber guarantees that this delay is larger than the coherence length of the light, the intensity at the end of the fiber is the sum of the intensities of the light traveling along each axis: $I = \cos^2 \theta I_{\text{Slow}} + \sin^2\theta I_{\text{Fast}}$, where $\theta$ is the anlge of polarization with respect to the fiber's principle axis, controlled by the orientation of the half wave plate at room temperature. The two orthogonal lineraly polarized beams emerging from the fiber go through a quarter wave-plate which is aligned such that the incident intensity on the sample is $I_{\text{inc}} = \cos^2 \theta I_{\text{Left}} + \sin^2\theta I_{\text{Right}}$. Thus, by rotating the half wave-plate, we can continuously change the proportion of right and left ciruclarly polarized light, without having linear or eliptically polarized light incident on the sample.

Upon illumination, some features in $R_{xx}$ are smeared (Fig. \ref{fig-heating}b). The light-induced changes to $R_{xx}$ are not persistent; the transport before illumination was recovered when the light was removed (Fig. \ref{fig-nomemoryRxx}). The smearing of the features in $R_{xx}$ could be caused by heating due to the light. Comparing the light-induced features with the thermal broadening due to heating the sample with a heater (Fig. \ref{fig-heating}c,d) reveals that the broadening is similar across all doping, and that the smearing at 200 $\mu$W corresponds roughly to 4 K. This is in sharp contrast to previous comparisons of transport under illumination \cite{Battista2022}, which showed that even at an incident power of 1 nW the electron temperature increased by about 1 K. Our sample may benefit from enhanced electron-phonon coupling due to alignment with h-BN and interaction with the WSe$_2$ layer \cite{Trovatello2022}. While the overall features in $R_{xx}$ are broadened by illumination, there is very little photo-doping ($<0.02$ electrons per moir\'e unit cell), as indicated by the positions of the correlated insulator resistance peaks, which overlap with the "dark" state.

\begin{figure*}[ht!]
\centering
\includegraphics[width=2.0\columnwidth]{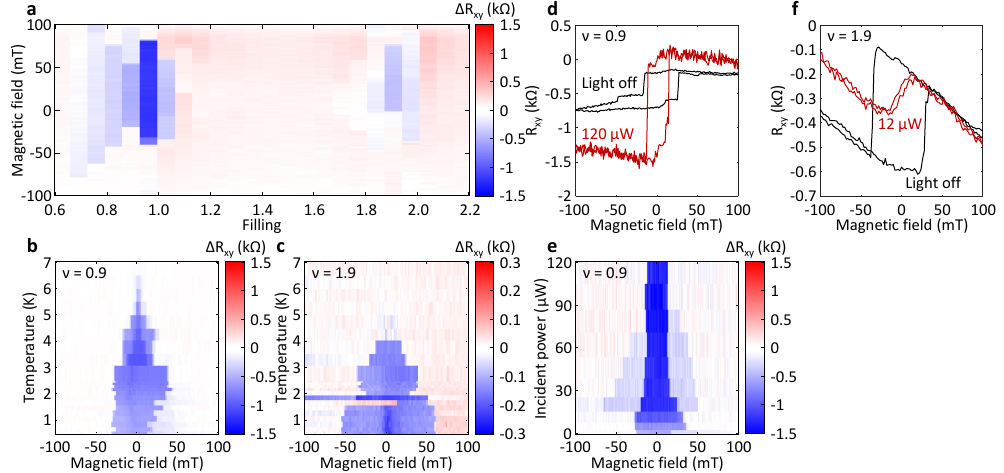}
\caption{The anomalous Hall effect at $\nu=1$. (a) The amplitude of the hysteresis loop, $\Delta R_{xy} = R_{xy}^{B\uparrow} - R_{xy}^{B\downarrow}$, plotted as a function of filling, showing an anomalous Hall effect near $\nu=1$ and $\nu=2$. The temperature dependence at $\nu = 0.9$ (b) and $\nu = 1.9$ (c) reveal Curie temperatures of 6 K and 4 K, respectively. (d) The hysteresis loop at $\nu = 0.9$ with and without illumination (120 $\mu$W, unpolarized light), showing a reduction in the coercive field along with an enhancement of the hysteresis amplitude. (e) $\Delta R_{xy}$ plotted for increasing light intensities at $\nu = 0.9$. (f) At $\nu = 1.9$, the hysteresis is eliminated upon illumination.}
\label{fig-lightAHE}
\end{figure*}

We turn now to the effect of light on the anomalous Hall states near $\nu=1$ and $\nu=2$. Figure \ref{fig-lightAHE}a shows $\Delta R_{xy} = R_{xy}^{B\uparrow} - R_{xy}^{B\downarrow}$, the difference in the Hall resistance between backward and forward sweeps of the magnetic field, as a function of filling. The data show typical hysteretic behavior at two filling bands, $0.7 \leq \nu \leq 1.1$ and $1.7 \leq \nu \leq 2$, with amplituds of about $1$ k$\Omega$ and a coercive field of about $50$ mT. The Curie temperatures were 6.5 K and 4 K respectively (Fig.\ref{fig-lightAHE}b,c). These observations are consistent with previous reports of orbital magnetism at odd \cite{Sharpe2019,Serlin2020} and even \cite{Stepanov2021,Tseng2022,Lin2022} integer fillings in MATBG, including with WSe$_2$ \cite{Bhowmik2023}.

Upon illuminating the sample near $\nu=1$ with an incident power of 120 $\mu$W,
the hysteresis in $R_{xy}$ increased in amplitude, while the coercive field decreased (Fig. \ref{fig-lightAHE}d). We measured the hysteresis as a function of the incident optical power (Fig. \ref{fig-lightAHE}e). At low powers, the hysteresis was similar to the dark behavior, with small jumps in the loop indicating the gradual switching of domains. Once the power exceeded 90 $\mu$W, we no longer observed small jumps and the coercive field and amplitude of the hysteresis where independent of the incident power, up to 200 $\mu$W. 
The reduction in coercive field is consistent with heating of the electronic system, as seen from the temperature dependence in Fig. \ref{fig-lightAHE}b. The increase in the hysteresis amplitude is similarly consistent, due to the temperature dependence of the longitudinal conductivity, $\sigma_{xx}$, which contributes to the Hall resistivity through $\rho_{xy}=-\sigma_{xy}/(\sigma_{xx}^2+\sigma_{xy}^2)$.

With $\sigma_{xx} \gg \sigma_{xy}$, the measured increase of $R_{xx}$ near $\nu=1$ upon warming or illumination (Fig. \ref{fig-photodoping}) indeed explains the observed  factor of $\sim$0.45 increase in $R_{xy}$.

Surprisingly, the AHE near $\nu=2$ is more susceptible to illumination despite its similar  Curie temperature.  Figure \ref{fig-lightAHE}f shows how the hysteresis is completely removed by illuminating with $12\, \mu$W. At this power, the broadening of $R_{xx}$ corresponds to a temperature increase well below the Curie temperature, suggesting that heating alone cannot explain the reduction of hysteresis upon illumination.
There are several possible explanations for the suppression of the AHE at $\nu = 2$. Firstly, noting the temperature dependence of $R_{xx}$, the $\nu = 2$ state shows stronger insulating behavior. As a result, the photo-induced carriers may have longer relaxation times \cite{Xie2024}, leading to an erasure of the orbital magentization during the relaxation. Alternatively, the differences between the AHE at $\nu = 1$ and $\nu = 2$ may point to different origins of the orbtial magnetization.  While the AHE at $\nu=1$ emerges from having one filled spin and valley polarized band, there are several possible configurations for the filling of the bands near $\nu=2$. Hartree-Fock calculations reveal closely competing intervalley-coherent states, intervalley Kekule spiral (IKS) and valley-polarized spin-unpolarized states \cite{Wagner2022,Breio2023}. The latter may be only partially polarized \cite{Bultnick2020}. While spin-orbit coupling induced by the WSe$_2$ in our sample favors a valley-polarized state, it is possible that the light promotes the IKS state, thereby suppressing the AHE. Alternatively, if the initial AHE is only partially polarized, electrons excited by the light can relax back into either valleys, thereby changing the overall magnetization.

\subsection{Inverse Faraday effect}
\begin{figure*}[ht!]
\includegraphics[width=2.0\columnwidth]{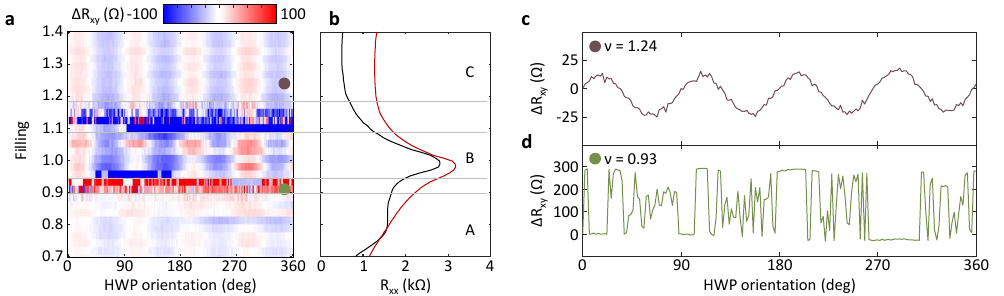}
\caption{Optical interrogation of the AHE. (a) Oscillations in $R_{xy}$ as a function of the half wave-plate orientation, tuning the incident polarization from left to right circularly polarized light. For some fillings close to $\nu =1$, we observe large jumps in $R_{xy}$ that are not periodic in the wave-plate orientation. (b) $R_{xx}$ as a function of filling at the corresponding range of oscillations. The fluctuations of $R_{xy}$ seen in panel a at specific fillings near $\nu = 0.95$ and $\nu = 1.1$ divide the phase diagram into three regimes, A with $\nu < 0.9$, B with $0.95 < \nu  < 1.15$ and C with $\nu > 1.15$. These regions are indicated by the gray lines. (c,d) Traces of the data in (a ) for $\nu = 1.24$ (c) and $\nu = 0.93$ (d).}
\label{fig-AHE_Pol}
\end{figure*}

Establishing the effect of light intensity, we turn now to explore the interaction of polarized light with the magnetic states near $\nu=1$ filling.  The common underlying mechanism to generate and manipulate magnetism in a medium with a finite magnetic susceptibility is the  inverse Faraday effect (IFE)  \cite{Pitaevskii1960,Ziel1965,Pershan1966}, where the magnitude and direction of magnetization in the material is modulated proportional to the difference in light intensities $( I_R - I_L )$ between right ($I_R$) and left ($I_L$), circularly polarized light. 
Thus, to test for the response of MATBG to circularly polarized light we measured $R_{xy}$ at zero magnetic field, while rotating the room-temperature half wave plate (HWP), thereby changing the ratio between right and left circularly polarized light created after passage through the quarter-waveplate (QWP)(Fig.~\ref{fig-heating}a). Fig.~\ref{fig-AHE_Pol}a shows the variations in $R_{xy}$ with  HWP orientation for various fillings, with the incident power fixed at 200 $\mu$W. For most fillings around $\nu=1$, $R_{xy}$ is periodically modulated as a function of the wave-plate orientation, with a period of 90\textdegree. The 90\textdegree\, period is set by the combination of the half wave-plate and the quarter wave-plate, which results in modulating the polarization from an equal intensity of the two circular polarizations to a pure circular polarization within 22.5\textdegree. The change, however, is small ($\sim 50$ \,$\Omega$) compared to the full range of magnetic hysteresis in $R_{xy}$ ($\sim 1$ k$\Omega$), suggesting that the light intensity we used produced a relatively weak effective magnetic field. Remarkably, at two narrow ranges of doping at $\nu = 0.95$ and $\nu=1.1$, (see Fig.~\ref{fig-AHE_Pol}a)  much larger switching events corresponding to $20-50\%$ of the full hysteresis of $\Delta R_{xy}$ is observed, and they occur without a clear periodicity with the incident polarization (Fig. \ref{fig-AHE_Pol}d), suggesting that light-induced periodic oscillations and random switching effects have different origins.

We note that the periodic oscillations occur in three distinct filling regimes, that are divided by the narrow filling bands where non-periodic, stronger in magnitude, fluctuations are observed: which we label A ($\nu < 0.9$), B ($0.95 < \nu  < 1.15$) and C ($\nu > 1.15$). Comparing to the observation of AHE in Fig.~\ref{fig-lightAHE}a, periodic oscillations in regions A (weak AHE) and region B (strong AHE) are observed on top of a remanent $R_{xy}$ and are strongly susceptible to emergence of non-periodic fluctuations if minor hysteresis loops are performed during illumination of the sample. Region C does not show a discernible hysteretic AHE, and by $\nu=1.3$ the Hall resistance has a straight linear dependence in field  (Fig.~\ref{fig-lightAHE}a). Therefore, this region exhibits periodic oscillations that are purely associated with the application of light. Furthermore, within the AHE region, oscillations were observed as long as $R_{xy}$ was not completely saturated. Applying a magnetic field to saturate $R_{xy}$ also removed the oscillations completely (Fig. \ref{fig-S_HWP_hyst}).

The oscillatory response may therefore be explained as a response of a magnetic system in the paramagnetic as well as the ordered regimes through the expression $\sigma_{xy}=\alpha M_{\text{eff}}$, where $\alpha$ is a proportionality constant. $M_{\text{eff}} = M_{\text{rem}} + M_{\text{ind}}$ is the effective magnetization, which is the sum of the remanent magnetization $M_{\text{rem}}$ and the optically induced magnetization $M_{\text{ind}}$. While $M_{\text{rem}}$ is a result of an external magnetic field or the anomalous Hall magnetization of the sample, $M_{\text{ind}}$, which depends on the value of the relevant susceptibility component, is directly related to the Verdet constant \cite{Pershan1966}. The observed values suggest that the susceptibility is much larger than that expected for electrons in the flat band (see Supplementary Information for discussion). At the same time, it is clear that the effect is strongly tied to the ferromagnetic state at $\nu=1$ as it is not observed in at fillings far away from the AHE regime (Fig. \ref{fig-no_osc_03}).

Where circularly polarized light is used, inverse magneto-optic effects are expected, which will induce a finite magnetization in the material via the inverse Faraday effect  \cite{Ziel1965,Pershan1966}. Relevant to our study is the induction of orbital magnetization through illumination with circularly polarized light \cite{Pershoguba2022,Sharma2024} recently observed in gold nanoparticles \cite{Cheng2020}  and the possible circularly-polarized light induced switching of orbital magnetization in the magnetized state \cite{Pershoguba2022}, recently proposed for TMD structures (see also perspective in \cite{Jimenez2024}). Since in region C, $R_{xx}$ is low compared to the peak magnetic state, we may analyze this regime as a IFE response of a two-dimensional conductor, focusing on the MATBG flat band.  Consider a two-dimensional isotropic medium of thickness $d$ in the $x$-$y$ plane with light of frequency $\omega$ impinging on it in the $+z$ direction. The Faraday angle, which quantifies the rotation of the plane of polarization of linearly polarized light as a result of a magnetic induction $B$ in the material is given by \cite{Argyres1955}
\begin{equation}
\theta_F=VBd\approx \frac{2\pi d}{c}\Re\Big\{\frac{\sigma_{xy}(\textbf{B})}{(1+4\pi i\sigma_{xx}/\omega)^{1/2}}\Big\}\approx\frac{2\pi d}{cn_0}\sigma_{xy}(\textbf{B}),
\label{argyres}
\end{equation}
where in the first equality we define the Verdet constant-$V$ and the last approximation refers to negligible dissipation in the materials and $n_0$ is the average index of refraction.

The inverse Faraday effect (IFE), where circularly polarized light induces a finite magnetization in the material is given by \cite{Pitaevskii1960,LL1984,Ziel1965,Pershan1966,Hertel2006}:
\begin{equation}
M_{\rm ind}=-\frac{cn_0}{8\pi \omega}V(i\textbf{E}\times\textbf{E}^*)=\frac{\lambda_0}{2\pi c}VI_C
\label{eqn-ife}
\end{equation}
where  in the second equality $I_C$ is the power of circularly polarized light per unit area incident on the material. To test the response of the MATBG system to circularly polarized light we measured its Faraday effect in the AHE state (Fig. \ref{fig_KF}). Since the MATBG is very thin (thickness $d=2d_g$, where $d_g\approx 2.6$~\AA~ \cite{Rickhaus2020} is the graphene dielectric thickness), the nominally Kerr experiment that we do is effectively a double Faraday effect as the beam returns from the substrate in our experiment acquires $2\times (2\theta_F)$ phase shift. 

 In the AHE state and at zero applied magnetic field, we can measure the Verdet constant directly by measuring the Faraday effect when the magnetization is saturated. While it is difficult to measure the saturation magnetization directly, it is expected that in a fully polarized AHE state it will correspond to $1\times\mu_B$ per moire unit cell $A_m\approx 100$ nm$^{2}$ \cite{Tschirhart2021,Grover2022}, resulting in $M_{\text{sat}} =\mu_B/(2d_g\times A_m) \approx 0.2$ G.  To find the Verdet constant, we measured the saturated Kerr/Faraday effect of the MATBG (see Supplemental Material). We find that at $\nu=0.85$, $R_{xy}$ saturation from positive to negative field is equivalent to a Faraday angle of $\theta_F\approx 10 \ \mu$rad. This in turn implies a Verdet constant of $V=\theta_F/M_{sat}2d_g\approx 1\times10^{-4} \ {\rm rad/cmOe}$. Returning to the measurement of $R_{xy}$ oscillations in Region C, we use the above Verdet constant in Eqn~\ref{eqn-ife} for  induced magnetization for light of intensity 120 $\mu$W at a spot size diameter of $\sim4.5 \ \mu$m. We obtain $\approx 0.05$ G, which is $\sim 0.025 M_{sat}$. Therefore, if $M_{sat}$ yielded anomalous Hall resistance of $\sim 1 \ k\Omega$, and only half the beam area was active, then  we expect $\sim 50 \ \Omega$ modulation, which is the value we observed experimentally. However, an estimation of the Verdet constant at 1550 nm from a semi-classical model results in a discrepancy of several orders of magnitudes, suggesting that the large value observed in the experiment is the result of a resonant excitation.

\subsection{Light-induced dynamics and percolation transition}
\begin{figure*}[ht!]
\includegraphics[width=2.0\columnwidth]{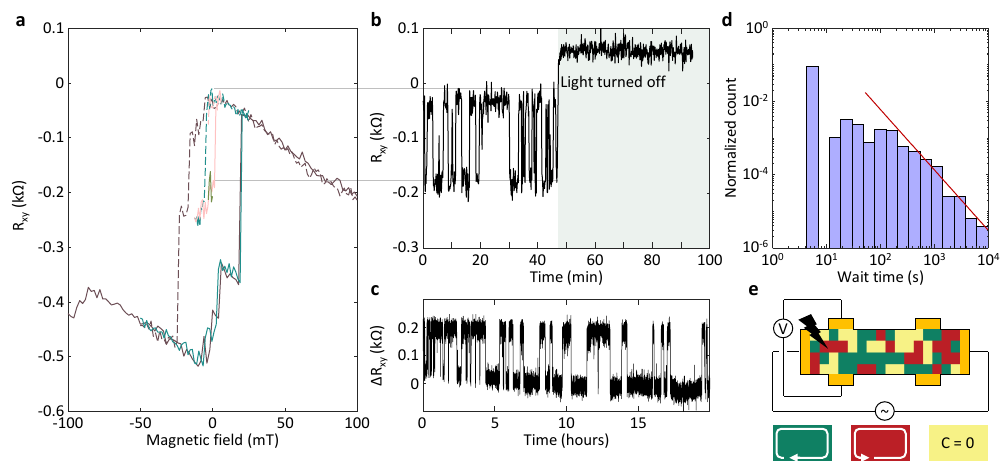}
\caption{Light-induced switching dynamics. (a) Major and minor hysteresis loops taken while illuminating the sample, at $\nu = 0.85$. The time-trace in panel (b) was taken at zero field, after performing a minor loop sweep from -12 mT. (b) Time fluctuations of $R_{xy}$ while the sample was illuminated. After the light was turned off the fluctuations stopped. The half waveplate orientation was kept fixed during this measurement. (c) Time fluctuations of $R_{xy}$ recorded at $\nu = 0.95$ over a 20-hour period. (d) A histogram of the wait times between switching events obtained from the time trace in (c). The data were binned into logarithmically spaced bins and the counts are normalized by the bin width and total number of events. The distribution follows a power law behavior, with an exponent $\approx 1.8$.
(e) Illustration of domains of opposite orbital magnetizations in an unmagnetized device. Near $\nu=1$ and a zero field, the device consists of a mixture of domains of opposite orbital magnetization, and therefore opposite edge currents, and trivial regions. The example illustrates how light-induced switching a single domain would generate a percolating path of domains between the transverse contacts, resulting in a large change to $R_{xy}$.}
\label{fig-percolation}
\end{figure*}

 In contrast to the oscillatory behavior observed at most fillings, at specific fillings close to $\nu=1$, we observed multiple sharp switching events that were only weakly correlated with the polarization of the light (Fig. \ref{fig-AHE_Pol}d). At these concentrations, we also observed sharp switching events over time, while keeping the orientation of the half wave-plate fixed (Fig. \ref{fig-percolation}a,b). When the light was turned off, we did not observe such switching. Remarkably, the switching occurs between two values which correspond to the size of a minor hysteresis loop that was performed before taking the time-dependent data. This suggests that the switching of $R_{xy}$ is directly related to the switching of domains of orbital magnetization. To further explore the switching dynamics, we recorded the switching over a 20 hour period. We observed periods with frequent switching events (a few seconds between switches), as well as long periods (10s of minutes) without any switching (Fig. \ref{fig-percolation}c). The distribution of wait-times between switching events (Fig. \ref{fig-percolation}d) obeys a power law dependence. This behavior is characteristic of systems without a well-defined time-scale, such as blinking quantum dots \cite{Stefani2009} and nano-magnetic particles \cite{Adhikari2024}.

The emerging picture is that the light-induced dynamics is driven by switching of single, or a small amount of domains, due to the weak interaction of the light with the correlated states. The light acts as a random field, for example due to the different interaction strength at different twist angle. The system can therefore be described using a random field Ising model and its fluctuations correspond to thermal switching of finite-size domains. Importantly, the resistance fluctuations in this model are largest when the system is close to its percolation threshold, meaning a patch of domains of one magnetization exists between the voltage contacts (Fig. \ref{fig-percolation}e). Thermal switching of domains along the percolating patch then results in large changes to $R_{xy}$. The switching rate strongly depends on the domain size \cite{Fisher1986}, resulting in multiple time-scales and therefore in a power-law spectrum, as we observed. The switching spectra of ferromagnets has been studied in the context of Barkhausen noise in ferromagnets\cite{Bittel1969}, and the resistance fluctuations have been analysed for a similar model describing the nematic phase transition of the cuprates \cite{Carlson2006}, where telegraph noise was observed in the resistance of nanowires in the pseudogap phase \cite{Bonetti2004}.

Indeed, the specific fillings at which the telegraph noise was observed correspond to two transitions in the system. The noise at filling $\nu \approx 1.1$ corresponds to the termination of the AHE regime, as indicated by the absence of hysteresis at higher fillings (Fig. \ref{fig-lightAHE}a), whereas the noise at $\nu \approx 0.9$ corresponds to the end of the "hump" region in $R_{xx}$, and the beggining of the sharp correlated insulator peak at $\nu = 1$. We note that the hump feature is caused by gapping out one of the Fermi surfaces (as indicated by its field dependence, see Fig. \ref{fig-photodoping}d) and its appearance coincides with the beginning of hysteresis in $R_{xy}$. We therefore interpret this hump region as a partially-valley-polarized state, which precedes the fully polarized state that occurs only at a narrow filling range around $\nu = 1$. The appearance of telegraph noise just before and just after the fully polarized state therefore indicates that the transition between the valley-polarized and the metallic states as a function of filling occurs via percolation.

\section{Conclusions}
To conclude, we studied the effects of near infrared light on the anomalous Hall state in MATBG. We revealed that the AHE is robust against light-induced heating. The survival of the AHE even under substantial optical power suggests that this phase, and possibly other stronly-correlated states in MATBG can be studied optically, for example via measurements of linear birefringence, circular dichroism and the polar Kerr effect, which will provide insights into the symmetry properties of the emergent phases. Furthermore, we demonstrated optical control over the orbital magnetism in MATBG, and uncovered the dynamics of the magnetic domains. We postulate that the fidelity of switching could be substantially increased by using terahertz radiation, below the gap between the flat and remote bands. Pulsed light could be used to provide ultra-fast control, opening a new avenue for engineering functional devices from van der Waals heterostructures, as well as high bandwidth photodetectors that are based on the large, digitized change to $R_{xy}$, in addition to bolometric detectors \cite{Deng2020,DiBattista2024}. Finally, the light-induced time dynamics are consistent with a percolation of magnetic domains. The absence of a characteristic time-scale for switching could provide insights into the energetics of domains in orbital magnets.

\section{Acknowledgments}
We thank Steve Kivelson, Julian May-Mann and Frank Koppens for fruitful discussions. Work at Stanford University was supported by  the U.~S.~Department of Energy (DOE) Office of Basic Energy Science, Division of Materials Science and Engineering at Stanford under contract No.~DE-AC02-76SF00515. EP was partially supported by the Koret Foundation. Work at the University of Washington is supported by NSF MRSEC DMR-1719797.

\bibliography{References.bib}

\begin{thebibliography}{54}%
\makeatletter
\providecommand \@ifxundefined [1]{%
 \@ifx{#1\undefined}
}%
\providecommand \@ifnum [1]{%
 \ifnum #1\expandafter \@firstoftwo
 \else \expandafter \@secondoftwo
 \fi
}%
\providecommand \@ifx [1]{%
 \ifx #1\expandafter \@firstoftwo
 \else \expandafter \@secondoftwo
 \fi
}%
\providecommand \natexlab [1]{#1}%
\providecommand \enquote  [1]{``#1''}%
\providecommand \bibnamefont  [1]{#1}%
\providecommand \bibfnamefont [1]{#1}%
\providecommand \citenamefont [1]{#1}%
\providecommand \href@noop [0]{\@secondoftwo}%
\providecommand \href [0]{\begingroup \@sanitize@url \@href}%
\providecommand \@href[1]{\@@startlink{#1}\@@href}%
\providecommand \@@href[1]{\endgroup#1\@@endlink}%
\providecommand \@sanitize@url [0]{\catcode `\\12\catcode `\$12\catcode `\&12\catcode `\#12\catcode `\^12\catcode `\_12\catcode `\%12\relax}%
\providecommand \@@startlink[1]{}%
\providecommand \@@endlink[0]{}%
\providecommand \url  [0]{\begingroup\@sanitize@url \@url }%
\providecommand \@url [1]{\endgroup\@href {#1}{\urlprefix }}%
\providecommand \urlprefix  [0]{URL }%
\providecommand \Eprint [0]{\href }%
\providecommand \doibase [0]{https://doi.org/}%
\providecommand \selectlanguage [0]{\@gobble}%
\providecommand \bibinfo  [0]{\@secondoftwo}%
\providecommand \bibfield  [0]{\@secondoftwo}%
\providecommand \translation [1]{[#1]}%
\providecommand \BibitemOpen [0]{}%
\providecommand \bibitemStop [0]{}%
\providecommand \bibitemNoStop [0]{.\EOS\space}%
\providecommand \EOS [0]{\spacefactor3000\relax}%
\providecommand \BibitemShut  [1]{\csname bibitem#1\endcsname}%
\let\auto@bib@innerbib\@empty
\bibitem [{\citenamefont {Cao}\ \emph {et~al.}(2018{\natexlab{a}})\citenamefont {Cao}, \citenamefont {Fatemi}, \citenamefont {Fang}, \citenamefont {Watanabe}, \citenamefont {Taniguchi}, \citenamefont {Kaxiras},\ and\ \citenamefont {Jarillo-Herrero}}]{Cao2018a}%
  \BibitemOpen
  \bibfield  {author} {\bibinfo {author} {\bibfnamefont {Y.}~\bibnamefont {Cao}}, \bibinfo {author} {\bibfnamefont {V.}~\bibnamefont {Fatemi}}, \bibinfo {author} {\bibfnamefont {S.}~\bibnamefont {Fang}}, \bibinfo {author} {\bibfnamefont {K.}~\bibnamefont {Watanabe}}, \bibinfo {author} {\bibfnamefont {T.}~\bibnamefont {Taniguchi}}, \bibinfo {author} {\bibfnamefont {E.}~\bibnamefont {Kaxiras}},\ and\ \bibinfo {author} {\bibfnamefont {P.}~\bibnamefont {Jarillo-Herrero}},\ }\bibfield  {title} {\bibinfo {title} {Unconventional superconductivity in magic-angle graphene superlattices},\ }\href {https://doi.org/10.1038/nature26160} {\bibfield  {journal} {\bibinfo  {journal} {Nature}\ }\textbf {\bibinfo {volume} {556}},\ \bibinfo {pages} {43–50} (\bibinfo {year} {2018}{\natexlab{a}})}\BibitemShut {NoStop}%
\bibitem [{\citenamefont {Cao}\ \emph {et~al.}(2018{\natexlab{b}})\citenamefont {Cao}, \citenamefont {Fatemi}, \citenamefont {Demir}, \citenamefont {Fang}, \citenamefont {Tomarken}, \citenamefont {Luo}, \citenamefont {Sanchez-Yamagishi}, \citenamefont {Watanabe}, \citenamefont {Taniguchi}, \citenamefont {Kaxiras}, \citenamefont {Ashoori},\ and\ \citenamefont {Jarillo-Herrero}}]{Cao2018b}%
  \BibitemOpen
  \bibfield  {author} {\bibinfo {author} {\bibfnamefont {Y.}~\bibnamefont {Cao}}, \bibinfo {author} {\bibfnamefont {V.}~\bibnamefont {Fatemi}}, \bibinfo {author} {\bibfnamefont {A.}~\bibnamefont {Demir}}, \bibinfo {author} {\bibfnamefont {S.}~\bibnamefont {Fang}}, \bibinfo {author} {\bibfnamefont {S.~L.}\ \bibnamefont {Tomarken}}, \bibinfo {author} {\bibfnamefont {J.~Y.}\ \bibnamefont {Luo}}, \bibinfo {author} {\bibfnamefont {J.~D.}\ \bibnamefont {Sanchez-Yamagishi}}, \bibinfo {author} {\bibfnamefont {K.}~\bibnamefont {Watanabe}}, \bibinfo {author} {\bibfnamefont {T.}~\bibnamefont {Taniguchi}}, \bibinfo {author} {\bibfnamefont {E.}~\bibnamefont {Kaxiras}}, \bibinfo {author} {\bibfnamefont {R.~C.}\ \bibnamefont {Ashoori}},\ and\ \bibinfo {author} {\bibfnamefont {P.}~\bibnamefont {Jarillo-Herrero}},\ }\bibfield  {title} {\bibinfo {title} {Correlated insulator behaviour at half-filling in magic-angle graphene superlattices},\ }\href {https://doi.org/10.1038/nature26154} {\bibfield  {journal} {\bibinfo  {journal}
  {Nature}\ }\textbf {\bibinfo {volume} {556}},\ \bibinfo {pages} {80–84} (\bibinfo {year} {2018}{\natexlab{b}})}\BibitemShut {NoStop}%
\bibitem [{\citenamefont {Sharpe}\ \emph {et~al.}(2019)\citenamefont {Sharpe}, \citenamefont {Fox}, \citenamefont {Barnard}, \citenamefont {Finney}, \citenamefont {Watanabe}, \citenamefont {Taniguchi}, \citenamefont {Kastner},\ and\ \citenamefont {Goldhaber-Gordon}}]{Sharpe2019}%
  \BibitemOpen
  \bibfield  {author} {\bibinfo {author} {\bibfnamefont {A.~L.}\ \bibnamefont {Sharpe}}, \bibinfo {author} {\bibfnamefont {E.~J.}\ \bibnamefont {Fox}}, \bibinfo {author} {\bibfnamefont {A.~W.}\ \bibnamefont {Barnard}}, \bibinfo {author} {\bibfnamefont {J.}~\bibnamefont {Finney}}, \bibinfo {author} {\bibfnamefont {K.}~\bibnamefont {Watanabe}}, \bibinfo {author} {\bibfnamefont {T.}~\bibnamefont {Taniguchi}}, \bibinfo {author} {\bibfnamefont {M.~A.}\ \bibnamefont {Kastner}},\ and\ \bibinfo {author} {\bibfnamefont {D.}~\bibnamefont {Goldhaber-Gordon}},\ }\bibfield  {title} {\bibinfo {title} {Emergent ferromagnetism near three-quarters filling in twisted bilayer graphene},\ }\href {https://doi.org/10.1126/science.aaw3780} {\bibfield  {journal} {\bibinfo  {journal} {Science}\ }\textbf {\bibinfo {volume} {365}},\ \bibinfo {pages} {605–608} (\bibinfo {year} {2019})}\BibitemShut {NoStop}%
\bibitem [{\citenamefont {Serlin}\ \emph {et~al.}(2020)\citenamefont {Serlin}, \citenamefont {Tschirhart}, \citenamefont {Polshyn}, \citenamefont {Zhang}, \citenamefont {Zhu}, \citenamefont {Watanabe}, \citenamefont {Taniguchi}, \citenamefont {Balents},\ and\ \citenamefont {Young}}]{Serlin2020}%
  \BibitemOpen
  \bibfield  {author} {\bibinfo {author} {\bibfnamefont {M.}~\bibnamefont {Serlin}}, \bibinfo {author} {\bibfnamefont {C.~L.}\ \bibnamefont {Tschirhart}}, \bibinfo {author} {\bibfnamefont {H.}~\bibnamefont {Polshyn}}, \bibinfo {author} {\bibfnamefont {Y.}~\bibnamefont {Zhang}}, \bibinfo {author} {\bibfnamefont {J.}~\bibnamefont {Zhu}}, \bibinfo {author} {\bibfnamefont {K.}~\bibnamefont {Watanabe}}, \bibinfo {author} {\bibfnamefont {T.}~\bibnamefont {Taniguchi}}, \bibinfo {author} {\bibfnamefont {L.}~\bibnamefont {Balents}},\ and\ \bibinfo {author} {\bibfnamefont {A.~F.}\ \bibnamefont {Young}},\ }\bibfield  {title} {\bibinfo {title} {Intrinsic quantized anomalous hall effect in a moiré heterostructure},\ }\href {https://doi.org/10.1126/science.aay5533} {\bibfield  {journal} {\bibinfo  {journal} {Science}\ }\textbf {\bibinfo {volume} {367}},\ \bibinfo {pages} {900–903} (\bibinfo {year} {2020})}\BibitemShut {NoStop}%
\bibitem [{\citenamefont {Xie}\ \emph {et~al.}(2021)\citenamefont {Xie}, \citenamefont {Pierce}, \citenamefont {Park}, \citenamefont {Parker}, \citenamefont {Khalaf}, \citenamefont {Ledwith}, \citenamefont {Cao}, \citenamefont {Lee}, \citenamefont {Chen}, \citenamefont {Forrester}, \citenamefont {Watanabe}, \citenamefont {Taniguchi}, \citenamefont {Vishwanath}, \citenamefont {Jarillo-Herrero},\ and\ \citenamefont {Yacoby}}]{Xie2021}%
  \BibitemOpen
  \bibfield  {author} {\bibinfo {author} {\bibfnamefont {Y.}~\bibnamefont {Xie}}, \bibinfo {author} {\bibfnamefont {A.~T.}\ \bibnamefont {Pierce}}, \bibinfo {author} {\bibfnamefont {J.~M.}\ \bibnamefont {Park}}, \bibinfo {author} {\bibfnamefont {D.~E.}\ \bibnamefont {Parker}}, \bibinfo {author} {\bibfnamefont {E.}~\bibnamefont {Khalaf}}, \bibinfo {author} {\bibfnamefont {P.}~\bibnamefont {Ledwith}}, \bibinfo {author} {\bibfnamefont {Y.}~\bibnamefont {Cao}}, \bibinfo {author} {\bibfnamefont {S.~H.}\ \bibnamefont {Lee}}, \bibinfo {author} {\bibfnamefont {S.}~\bibnamefont {Chen}}, \bibinfo {author} {\bibfnamefont {P.~R.}\ \bibnamefont {Forrester}}, \bibinfo {author} {\bibfnamefont {K.}~\bibnamefont {Watanabe}}, \bibinfo {author} {\bibfnamefont {T.}~\bibnamefont {Taniguchi}}, \bibinfo {author} {\bibfnamefont {A.}~\bibnamefont {Vishwanath}}, \bibinfo {author} {\bibfnamefont {P.}~\bibnamefont {Jarillo-Herrero}},\ and\ \bibinfo {author} {\bibfnamefont {A.}~\bibnamefont {Yacoby}},\ }\bibfield  {title} {\bibinfo
  {title} {Fractional chern insulators in magic-angle twisted bilayer graphene},\ }\href {https://doi.org/10.1038/s41586-021-04002-3} {\bibfield  {journal} {\bibinfo  {journal} {Nature}\ }\textbf {\bibinfo {volume} {600}},\ \bibinfo {pages} {439–443} (\bibinfo {year} {2021})}\BibitemShut {NoStop}%
\bibitem [{\citenamefont {Jiang}\ \emph {et~al.}(2019)\citenamefont {Jiang}, \citenamefont {Lai}, \citenamefont {Watanabe}, \citenamefont {Taniguchi}, \citenamefont {Haule}, \citenamefont {Mao},\ and\ \citenamefont {Andrei}}]{Jiang2019}%
  \BibitemOpen
  \bibfield  {author} {\bibinfo {author} {\bibfnamefont {Y.}~\bibnamefont {Jiang}}, \bibinfo {author} {\bibfnamefont {X.}~\bibnamefont {Lai}}, \bibinfo {author} {\bibfnamefont {K.}~\bibnamefont {Watanabe}}, \bibinfo {author} {\bibfnamefont {T.}~\bibnamefont {Taniguchi}}, \bibinfo {author} {\bibfnamefont {K.}~\bibnamefont {Haule}}, \bibinfo {author} {\bibfnamefont {J.}~\bibnamefont {Mao}},\ and\ \bibinfo {author} {\bibfnamefont {E.~Y.}\ \bibnamefont {Andrei}},\ }\bibfield  {title} {\bibinfo {title} {Charge order and broken rotational symmetry in magic-angle twisted bilayer graphene},\ }\href {https://doi.org/10.1038/s41586-019-1460-4} {\bibfield  {journal} {\bibinfo  {journal} {Nature}\ }\textbf {\bibinfo {volume} {573}},\ \bibinfo {pages} {91–95} (\bibinfo {year} {2019})}\BibitemShut {NoStop}%
\bibitem [{\citenamefont {Bistritzer}\ and\ \citenamefont {MacDonald}(2011)}]{Bistritzer2011}%
  \BibitemOpen
  \bibfield  {author} {\bibinfo {author} {\bibfnamefont {R.}~\bibnamefont {Bistritzer}}\ and\ \bibinfo {author} {\bibfnamefont {A.~H.}\ \bibnamefont {MacDonald}},\ }\bibfield  {title} {\bibinfo {title} {Moiré bands in twisted double-layer graphene},\ }\href {https://doi.org/10.1073/pnas.1108174108} {\bibfield  {journal} {\bibinfo  {journal} {Proceedings of the National Academy of Sciences}\ }\textbf {\bibinfo {volume} {108}},\ \bibinfo {pages} {12233} (\bibinfo {year} {2011})}\BibitemShut {NoStop}%
\bibitem [{\citenamefont {Polski}\ \emph {et~al.}(2022)\citenamefont {Polski}, \citenamefont {Zhang}, \citenamefont {Peng}, \citenamefont {Arora}, \citenamefont {Choi}, \citenamefont {Kim}, \citenamefont {Watanabe}, \citenamefont {Taniguchi}, \citenamefont {Refael}, \citenamefont {von Oppen},\ and\ \citenamefont {Nadj-Perge}}]{Polski2022}%
  \BibitemOpen
  \bibfield  {author} {\bibinfo {author} {\bibfnamefont {R.}~\bibnamefont {Polski}}, \bibinfo {author} {\bibfnamefont {Y.}~\bibnamefont {Zhang}}, \bibinfo {author} {\bibfnamefont {Y.}~\bibnamefont {Peng}}, \bibinfo {author} {\bibfnamefont {H.~S.}\ \bibnamefont {Arora}}, \bibinfo {author} {\bibfnamefont {Y.}~\bibnamefont {Choi}}, \bibinfo {author} {\bibfnamefont {H.}~\bibnamefont {Kim}}, \bibinfo {author} {\bibfnamefont {K.}~\bibnamefont {Watanabe}}, \bibinfo {author} {\bibfnamefont {T.}~\bibnamefont {Taniguchi}}, \bibinfo {author} {\bibfnamefont {G.}~\bibnamefont {Refael}}, \bibinfo {author} {\bibfnamefont {F.}~\bibnamefont {von Oppen}},\ and\ \bibinfo {author} {\bibfnamefont {S.}~\bibnamefont {Nadj-Perge}},\ }\href@noop {} {\bibinfo {title} {Hierarchy of symmetry breaking correlated phases in twisted bilayer graphene}} (\bibinfo {year} {2022}),\ \Eprint {https://arxiv.org/abs/arXiv:2205.05225} {arXiv:2205.05225} \BibitemShut {NoStop}%
\bibitem [{\citenamefont {He}\ \emph {et~al.}(2024)\citenamefont {He}, \citenamefont {Wang}, \citenamefont {Cai}, \citenamefont {Herzog-Arbeitman}, \citenamefont {Taniguchi}, \citenamefont {Watanabe}, \citenamefont {Stern}, \citenamefont {Bernevig}, \citenamefont {Yankowitz}, \citenamefont {Vafek},\ and\ \citenamefont {Xu}}]{He2024}%
  \BibitemOpen
  \bibfield  {author} {\bibinfo {author} {\bibfnamefont {M.}~\bibnamefont {He}}, \bibinfo {author} {\bibfnamefont {X.}~\bibnamefont {Wang}}, \bibinfo {author} {\bibfnamefont {J.}~\bibnamefont {Cai}}, \bibinfo {author} {\bibfnamefont {J.}~\bibnamefont {Herzog-Arbeitman}}, \bibinfo {author} {\bibfnamefont {T.}~\bibnamefont {Taniguchi}}, \bibinfo {author} {\bibfnamefont {K.}~\bibnamefont {Watanabe}}, \bibinfo {author} {\bibfnamefont {A.}~\bibnamefont {Stern}}, \bibinfo {author} {\bibfnamefont {B.~A.}\ \bibnamefont {Bernevig}}, \bibinfo {author} {\bibfnamefont {M.}~\bibnamefont {Yankowitz}}, \bibinfo {author} {\bibfnamefont {O.}~\bibnamefont {Vafek}},\ and\ \bibinfo {author} {\bibfnamefont {X.}~\bibnamefont {Xu}},\ }\href@noop {} {\bibinfo {title} {Strongly interacting hofstadter states in magic-angle twisted bilayer graphene}} (\bibinfo {year} {2024}),\ \Eprint {https://arxiv.org/abs/arXiv:2408.01599} {arXiv:2408.01599} \BibitemShut {NoStop}%
\bibitem [{\citenamefont {Liu}\ and\ \citenamefont {Dai}(2020)}]{Liu2020}%
  \BibitemOpen
  \bibfield  {author} {\bibinfo {author} {\bibfnamefont {J.}~\bibnamefont {Liu}}\ and\ \bibinfo {author} {\bibfnamefont {X.}~\bibnamefont {Dai}},\ }\bibfield  {title} {\bibinfo {title} {Anomalous hall effect, magneto-optical properties, and nonlinear optical properties of twisted graphene systems},\ }\href {https://doi.org/10.1038/s41524-020-0299-4} {\bibfield  {journal} {\bibinfo  {journal} {npj Computational Materials}\ }\textbf {\bibinfo {volume} {6}},\ \bibinfo {pages} {57} (\bibinfo {year} {2020})}\BibitemShut {NoStop}%
\bibitem [{\citenamefont {Liu}\ \emph {et~al.}(2024)\citenamefont {Liu}, \citenamefont {Liu}, \citenamefont {Cao},\ and\ \citenamefont {Wang}}]{Liu2024}%
  \BibitemOpen
  \bibfield  {author} {\bibinfo {author} {\bibfnamefont {M.}~\bibnamefont {Liu}}, \bibinfo {author} {\bibfnamefont {Z.}~\bibnamefont {Liu}}, \bibinfo {author} {\bibfnamefont {J.}~\bibnamefont {Cao}},\ and\ \bibinfo {author} {\bibfnamefont {C.}~\bibnamefont {Wang}},\ }\bibfield  {title} {\bibinfo {title} {Properties of the optical response of the twisted bilayer graphene},\ }\href {https://doi.org/https://doi.org/10.1016/j.physb.2023.415609} {\bibfield  {journal} {\bibinfo  {journal} {Physica B: Condensed Matter}\ }\textbf {\bibinfo {volume} {675}},\ \bibinfo {pages} {415609} (\bibinfo {year} {2024})}\BibitemShut {NoStop}%
\bibitem [{\citenamefont {Wu}\ \emph {et~al.}(2018)\citenamefont {Wu}, \citenamefont {Lovorn}, \citenamefont {Tutuc},\ and\ \citenamefont {MacDonald}}]{Wu2018}%
  \BibitemOpen
  \bibfield  {author} {\bibinfo {author} {\bibfnamefont {F.}~\bibnamefont {Wu}}, \bibinfo {author} {\bibfnamefont {T.}~\bibnamefont {Lovorn}}, \bibinfo {author} {\bibfnamefont {E.}~\bibnamefont {Tutuc}},\ and\ \bibinfo {author} {\bibfnamefont {A.~H.}\ \bibnamefont {MacDonald}},\ }\bibfield  {title} {\bibinfo {title} {Hubbard model physics in transition metal dichalcogenide moir\'e bands},\ }\href {https://doi.org/10.1103/PhysRevLett.121.026402} {\bibfield  {journal} {\bibinfo  {journal} {Phys. Rev. Lett.}\ }\textbf {\bibinfo {volume} {121}},\ \bibinfo {pages} {026402} (\bibinfo {year} {2018})}\BibitemShut {NoStop}%
\bibitem [{\citenamefont {Cai}\ \emph {et~al.}(2023)\citenamefont {Cai}, \citenamefont {Anderson}, \citenamefont {Wang}, \citenamefont {Zhang}, \citenamefont {Liu}, \citenamefont {Holtzmann}, \citenamefont {Zhang}, \citenamefont {Fan}, \citenamefont {Taniguchi}, \citenamefont {Watanabe}, \citenamefont {Ran}, \citenamefont {Cao}, \citenamefont {Fu}, \citenamefont {Xiao}, \citenamefont {Yao},\ and\ \citenamefont {Xu}}]{Cai2023}%
  \BibitemOpen
  \bibfield  {author} {\bibinfo {author} {\bibfnamefont {J.}~\bibnamefont {Cai}}, \bibinfo {author} {\bibfnamefont {E.}~\bibnamefont {Anderson}}, \bibinfo {author} {\bibfnamefont {C.}~\bibnamefont {Wang}}, \bibinfo {author} {\bibfnamefont {X.}~\bibnamefont {Zhang}}, \bibinfo {author} {\bibfnamefont {X.}~\bibnamefont {Liu}}, \bibinfo {author} {\bibfnamefont {W.}~\bibnamefont {Holtzmann}}, \bibinfo {author} {\bibfnamefont {Y.}~\bibnamefont {Zhang}}, \bibinfo {author} {\bibfnamefont {F.}~\bibnamefont {Fan}}, \bibinfo {author} {\bibfnamefont {T.}~\bibnamefont {Taniguchi}}, \bibinfo {author} {\bibfnamefont {K.}~\bibnamefont {Watanabe}}, \bibinfo {author} {\bibfnamefont {Y.}~\bibnamefont {Ran}}, \bibinfo {author} {\bibfnamefont {T.}~\bibnamefont {Cao}}, \bibinfo {author} {\bibfnamefont {L.}~\bibnamefont {Fu}}, \bibinfo {author} {\bibfnamefont {D.}~\bibnamefont {Xiao}}, \bibinfo {author} {\bibfnamefont {W.}~\bibnamefont {Yao}},\ and\ \bibinfo {author} {\bibfnamefont {X.}~\bibnamefont {Xu}},\ }\bibfield  {title}
  {\bibinfo {title} {Signatures of fractional quantum anomalous hall states in twisted $\textrm{MoTe}_2$},\ }\href {https://doi.org/10.1038/s41586-023-06289-w} {\bibfield  {journal} {\bibinfo  {journal} {Nature}\ }\textbf {\bibinfo {volume} {622}},\ \bibinfo {pages} {63–68} (\bibinfo {year} {2023})}\BibitemShut {NoStop}%
\bibitem [{\citenamefont {Zeng}\ \emph {et~al.}(2023)\citenamefont {Zeng}, \citenamefont {Xia}, \citenamefont {Kang}, \citenamefont {Zhu}, \citenamefont {Kn\"{u}ppel}, \citenamefont {Vaswani}, \citenamefont {Watanabe}, \citenamefont {Taniguchi}, \citenamefont {Mak},\ and\ \citenamefont {Shan}}]{Zeng2023}%
  \BibitemOpen
  \bibfield  {author} {\bibinfo {author} {\bibfnamefont {Y.}~\bibnamefont {Zeng}}, \bibinfo {author} {\bibfnamefont {Z.}~\bibnamefont {Xia}}, \bibinfo {author} {\bibfnamefont {K.}~\bibnamefont {Kang}}, \bibinfo {author} {\bibfnamefont {J.}~\bibnamefont {Zhu}}, \bibinfo {author} {\bibfnamefont {P.}~\bibnamefont {Kn\"{u}ppel}}, \bibinfo {author} {\bibfnamefont {C.}~\bibnamefont {Vaswani}}, \bibinfo {author} {\bibfnamefont {K.}~\bibnamefont {Watanabe}}, \bibinfo {author} {\bibfnamefont {T.}~\bibnamefont {Taniguchi}}, \bibinfo {author} {\bibfnamefont {K.~F.}\ \bibnamefont {Mak}},\ and\ \bibinfo {author} {\bibfnamefont {J.}~\bibnamefont {Shan}},\ }\bibfield  {title} {\bibinfo {title} {Thermodynamic evidence of fractional chern insulator in moiré $\textrm{MoTe}_2$},\ }\href {https://doi.org/10.1038/s41586-023-06452-3} {\bibfield  {journal} {\bibinfo  {journal} {Nature}\ }\textbf {\bibinfo {volume} {622}},\ \bibinfo {pages} {69–73} (\bibinfo {year} {2023})}\BibitemShut {NoStop}%
\bibitem [{\citenamefont {Chernikov}\ \emph {et~al.}(2014)\citenamefont {Chernikov}, \citenamefont {Berkelbach}, \citenamefont {Hill}, \citenamefont {Rigosi}, \citenamefont {Li}, \citenamefont {Aslan}, \citenamefont {Reichman}, \citenamefont {Hybertsen},\ and\ \citenamefont {Heinz}}]{Chernikov2014}%
  \BibitemOpen
  \bibfield  {author} {\bibinfo {author} {\bibfnamefont {A.}~\bibnamefont {Chernikov}}, \bibinfo {author} {\bibfnamefont {T.~C.}\ \bibnamefont {Berkelbach}}, \bibinfo {author} {\bibfnamefont {H.~M.}\ \bibnamefont {Hill}}, \bibinfo {author} {\bibfnamefont {A.}~\bibnamefont {Rigosi}}, \bibinfo {author} {\bibfnamefont {Y.}~\bibnamefont {Li}}, \bibinfo {author} {\bibfnamefont {B.}~\bibnamefont {Aslan}}, \bibinfo {author} {\bibfnamefont {D.~R.}\ \bibnamefont {Reichman}}, \bibinfo {author} {\bibfnamefont {M.~S.}\ \bibnamefont {Hybertsen}},\ and\ \bibinfo {author} {\bibfnamefont {T.~F.}\ \bibnamefont {Heinz}},\ }\bibfield  {title} {\bibinfo {title} {Exciton binding energy and nonhydrogenic rydberg series in monolayer $\textrm{WS}_{2}$},\ }\href {https://doi.org/10.1103/PhysRevLett.113.076802} {\bibfield  {journal} {\bibinfo  {journal} {Phys. Rev. Lett.}\ }\textbf {\bibinfo {volume} {113}},\ \bibinfo {pages} {076802} (\bibinfo {year} {2014})}\BibitemShut {NoStop}%
\bibitem [{\citenamefont {Tielrooij}\ \emph {et~al.}(2013)\citenamefont {Tielrooij}, \citenamefont {Song}, \citenamefont {Jensen}, \citenamefont {Centeno}, \citenamefont {Pesquera}, \citenamefont {Zurutuza~Elorza}, \citenamefont {Bonn}, \citenamefont {Levitov},\ and\ \citenamefont {Koppens}}]{Tielrooij2013}%
  \BibitemOpen
  \bibfield  {author} {\bibinfo {author} {\bibfnamefont {K.~J.}\ \bibnamefont {Tielrooij}}, \bibinfo {author} {\bibfnamefont {J.~C.~W.}\ \bibnamefont {Song}}, \bibinfo {author} {\bibfnamefont {S.~A.}\ \bibnamefont {Jensen}}, \bibinfo {author} {\bibfnamefont {A.}~\bibnamefont {Centeno}}, \bibinfo {author} {\bibfnamefont {A.}~\bibnamefont {Pesquera}}, \bibinfo {author} {\bibfnamefont {A.}~\bibnamefont {Zurutuza~Elorza}}, \bibinfo {author} {\bibfnamefont {M.}~\bibnamefont {Bonn}}, \bibinfo {author} {\bibfnamefont {L.~S.}\ \bibnamefont {Levitov}},\ and\ \bibinfo {author} {\bibfnamefont {F.~H.~L.}\ \bibnamefont {Koppens}},\ }\bibfield  {title} {\bibinfo {title} {Photoexcitation cascade and multiple hot-carrier generation in graphene},\ }\href {https://doi.org/10.1038/nphys2564} {\bibfield  {journal} {\bibinfo  {journal} {Nature Physics}\ }\textbf {\bibinfo {volume} {9}},\ \bibinfo {pages} {248–252} (\bibinfo {year} {2013})}\BibitemShut {NoStop}%
\bibitem [{\citenamefont {Gierz}\ \emph {et~al.}(2013)\citenamefont {Gierz}, \citenamefont {Petersen}, \citenamefont {Mitrano}, \citenamefont {Cacho}, \citenamefont {Turcu}, \citenamefont {Springate}, \citenamefont {St\"{o}hr}, \citenamefont {K\"{o}hler}, \citenamefont {Starke},\ and\ \citenamefont {Cavalleri}}]{Gierz2013}%
  \BibitemOpen
  \bibfield  {author} {\bibinfo {author} {\bibfnamefont {I.}~\bibnamefont {Gierz}}, \bibinfo {author} {\bibfnamefont {J.~C.}\ \bibnamefont {Petersen}}, \bibinfo {author} {\bibfnamefont {M.}~\bibnamefont {Mitrano}}, \bibinfo {author} {\bibfnamefont {C.}~\bibnamefont {Cacho}}, \bibinfo {author} {\bibfnamefont {I.~C.~E.}\ \bibnamefont {Turcu}}, \bibinfo {author} {\bibfnamefont {E.}~\bibnamefont {Springate}}, \bibinfo {author} {\bibfnamefont {A.}~\bibnamefont {St\"{o}hr}}, \bibinfo {author} {\bibfnamefont {A.}~\bibnamefont {K\"{o}hler}}, \bibinfo {author} {\bibfnamefont {U.}~\bibnamefont {Starke}},\ and\ \bibinfo {author} {\bibfnamefont {A.}~\bibnamefont {Cavalleri}},\ }\bibfield  {title} {\bibinfo {title} {Snapshots of non-equilibrium dirac carrier distributions in graphene},\ }\href {https://doi.org/10.1038/nmat3757} {\bibfield  {journal} {\bibinfo  {journal} {Nature Materials}\ }\textbf {\bibinfo {volume} {12}},\ \bibinfo {pages} {1119–1124} (\bibinfo {year} {2013})}\BibitemShut {NoStop}%
\bibitem [{\citenamefont {Di~Battista}\ \emph {et~al.}(2022)\citenamefont {Di~Battista}, \citenamefont {Seifert}, \citenamefont {Watanabe}, \citenamefont {Taniguchi}, \citenamefont {Fong}, \citenamefont {Principi},\ and\ \citenamefont {Efetov}}]{Battista2022}%
  \BibitemOpen
  \bibfield  {author} {\bibinfo {author} {\bibfnamefont {G.}~\bibnamefont {Di~Battista}}, \bibinfo {author} {\bibfnamefont {P.}~\bibnamefont {Seifert}}, \bibinfo {author} {\bibfnamefont {K.}~\bibnamefont {Watanabe}}, \bibinfo {author} {\bibfnamefont {T.}~\bibnamefont {Taniguchi}}, \bibinfo {author} {\bibfnamefont {K.~C.}\ \bibnamefont {Fong}}, \bibinfo {author} {\bibfnamefont {A.}~\bibnamefont {Principi}},\ and\ \bibinfo {author} {\bibfnamefont {D.~K.}\ \bibnamefont {Efetov}},\ }\bibfield  {title} {\bibinfo {title} {Revealing the thermal properties of superconducting magic-angle twisted bilayer graphene},\ }\href {https://doi.org/10.1021/acs.nanolett.1c04512} {\bibfield  {journal} {\bibinfo  {journal} {Nano Letters}\ }\textbf {\bibinfo {volume} {22}},\ \bibinfo {pages} {6465} (\bibinfo {year} {2022})}\BibitemShut {NoStop}%
\bibitem [{\citenamefont {Merino}\ \emph {et~al.}(2024)\citenamefont {Merino}, \citenamefont {Calugaru}, \citenamefont {Hu}, \citenamefont {Diez-Merida}, \citenamefont {Diez-Carlon}, \citenamefont {Taniguchi}, \citenamefont {Watanabe}, \citenamefont {Seifert}, \citenamefont {Bernevig},\ and\ \citenamefont {Efetov}}]{Merino2024}%
  \BibitemOpen
  \bibfield  {author} {\bibinfo {author} {\bibfnamefont {R.~L.}\ \bibnamefont {Merino}}, \bibinfo {author} {\bibfnamefont {D.}~\bibnamefont {Calugaru}}, \bibinfo {author} {\bibfnamefont {H.}~\bibnamefont {Hu}}, \bibinfo {author} {\bibfnamefont {J.}~\bibnamefont {Diez-Merida}}, \bibinfo {author} {\bibfnamefont {A.}~\bibnamefont {Diez-Carlon}}, \bibinfo {author} {\bibfnamefont {T.}~\bibnamefont {Taniguchi}}, \bibinfo {author} {\bibfnamefont {K.}~\bibnamefont {Watanabe}}, \bibinfo {author} {\bibfnamefont {P.}~\bibnamefont {Seifert}}, \bibinfo {author} {\bibfnamefont {B.~A.}\ \bibnamefont {Bernevig}},\ and\ \bibinfo {author} {\bibfnamefont {D.~K.}\ \bibnamefont {Efetov}},\ }\href@noop {} {\bibinfo {title} {Evidence of heavy fermion physics in the thermoelectric transport of magic angle twisted bilayer graphene}} (\bibinfo {year} {2024}),\ \Eprint {https://arxiv.org/abs/arXiv:2402.11749} {arXiv:2402.11749} \BibitemShut {NoStop}%
\bibitem [{\citenamefont {Pershoguba}\ and\ \citenamefont {Yakovenko}(2022)}]{Pershoguba2022}%
  \BibitemOpen
  \bibfield  {author} {\bibinfo {author} {\bibfnamefont {S.~S.}\ \bibnamefont {Pershoguba}}\ and\ \bibinfo {author} {\bibfnamefont {V.~M.}\ \bibnamefont {Yakovenko}},\ }\bibfield  {title} {\bibinfo {title} {Optical control of topological memory based on orbital magnetization},\ }\href {https://doi.org/10.1103/PhysRevB.105.064423} {\bibfield  {journal} {\bibinfo  {journal} {Phys. Rev. B}\ }\textbf {\bibinfo {volume} {105}},\ \bibinfo {pages} {064423} (\bibinfo {year} {2022})}\BibitemShut {NoStop}%
\bibitem [{\citenamefont {Yang}\ \emph {et~al.}(2023)\citenamefont {Yang}, \citenamefont {Esin}, \citenamefont {Lewandowski},\ and\ \citenamefont {Refael}}]{Yang2023}%
  \BibitemOpen
  \bibfield  {author} {\bibinfo {author} {\bibfnamefont {C.}~\bibnamefont {Yang}}, \bibinfo {author} {\bibfnamefont {I.}~\bibnamefont {Esin}}, \bibinfo {author} {\bibfnamefont {C.}~\bibnamefont {Lewandowski}},\ and\ \bibinfo {author} {\bibfnamefont {G.}~\bibnamefont {Refael}},\ }\bibfield  {title} {\bibinfo {title} {Optical control of slow topological electrons in moir\'e systems},\ }\href {https://doi.org/10.1103/PhysRevLett.131.026901} {\bibfield  {journal} {\bibinfo  {journal} {Phys. Rev. Lett.}\ }\textbf {\bibinfo {volume} {131}},\ \bibinfo {pages} {026901} (\bibinfo {year} {2023})}\BibitemShut {NoStop}%
\bibitem [{\citenamefont {Kumar}\ \emph {et~al.}(2024)\citenamefont {Kumar}, \citenamefont {Li}, \citenamefont {Bertini}, \citenamefont {Chaudhary}, \citenamefont {Nowakowski}, \citenamefont {Park}, \citenamefont {Castilla}, \citenamefont {Zhan}, \citenamefont {Pantaleón}, \citenamefont {Agarwal}, \citenamefont {Battle-Porro}, \citenamefont {Icking}, \citenamefont {Ceccanti}, \citenamefont {Reserbat-Plantey}, \citenamefont {Piccinini}, \citenamefont {Barrier}, \citenamefont {Khestanova}, \citenamefont {Taniguchi}, \citenamefont {Watanabe}, \citenamefont {Stampfer}, \citenamefont {Refael}, \citenamefont {Guinea}, \citenamefont {Jarillo-Herrero}, \citenamefont {Song}, \citenamefont {Stepanov}, \citenamefont {Lewandowski},\ and\ \citenamefont {Koppens}}]{Kumar2024}%
  \BibitemOpen
  \bibfield  {author} {\bibinfo {author} {\bibfnamefont {R.~K.}\ \bibnamefont {Kumar}}, \bibinfo {author} {\bibfnamefont {G.}~\bibnamefont {Li}}, \bibinfo {author} {\bibfnamefont {R.}~\bibnamefont {Bertini}}, \bibinfo {author} {\bibfnamefont {S.}~\bibnamefont {Chaudhary}}, \bibinfo {author} {\bibfnamefont {K.}~\bibnamefont {Nowakowski}}, \bibinfo {author} {\bibfnamefont {J.~M.}\ \bibnamefont {Park}}, \bibinfo {author} {\bibfnamefont {S.}~\bibnamefont {Castilla}}, \bibinfo {author} {\bibfnamefont {Z.}~\bibnamefont {Zhan}}, \bibinfo {author} {\bibfnamefont {P.~A.}\ \bibnamefont {Pantaleón}}, \bibinfo {author} {\bibfnamefont {H.}~\bibnamefont {Agarwal}}, \bibinfo {author} {\bibfnamefont {S.}~\bibnamefont {Battle-Porro}}, \bibinfo {author} {\bibfnamefont {E.}~\bibnamefont {Icking}}, \bibinfo {author} {\bibfnamefont {M.}~\bibnamefont {Ceccanti}}, \bibinfo {author} {\bibfnamefont {A.}~\bibnamefont {Reserbat-Plantey}}, \bibinfo {author} {\bibfnamefont {G.}~\bibnamefont {Piccinini}}, \bibinfo {author} {\bibfnamefont
  {J.}~\bibnamefont {Barrier}}, \bibinfo {author} {\bibfnamefont {E.}~\bibnamefont {Khestanova}}, \bibinfo {author} {\bibfnamefont {T.}~\bibnamefont {Taniguchi}}, \bibinfo {author} {\bibfnamefont {K.}~\bibnamefont {Watanabe}}, \bibinfo {author} {\bibfnamefont {C.}~\bibnamefont {Stampfer}}, \bibinfo {author} {\bibfnamefont {G.}~\bibnamefont {Refael}}, \bibinfo {author} {\bibfnamefont {F.}~\bibnamefont {Guinea}}, \bibinfo {author} {\bibfnamefont {P.}~\bibnamefont {Jarillo-Herrero}}, \bibinfo {author} {\bibfnamefont {J.~C.~W.}\ \bibnamefont {Song}}, \bibinfo {author} {\bibfnamefont {P.}~\bibnamefont {Stepanov}}, \bibinfo {author} {\bibfnamefont {C.}~\bibnamefont {Lewandowski}},\ and\ \bibinfo {author} {\bibfnamefont {F.~H.~L.}\ \bibnamefont {Koppens}},\ }\href@noop {} {\bibinfo {title} {Terahertz photocurrent probe of quantum geometry and interactions in magic-angle twisted bilayer graphene}} (\bibinfo {year} {2024}),\ \Eprint {https://arxiv.org/abs/arXiv:2406.16532} {arXiv:2406.16532} \BibitemShut {NoStop}%
\bibitem [{\citenamefont {Lin}\ \emph {et~al.}(2022)\citenamefont {Lin}, \citenamefont {Zhang}, \citenamefont {Morissette}, \citenamefont {Wang}, \citenamefont {Liu}, \citenamefont {Rhodes}, \citenamefont {Watanabe}, \citenamefont {Taniguchi}, \citenamefont {Hone},\ and\ \citenamefont {Li}}]{Lin2022}%
  \BibitemOpen
  \bibfield  {author} {\bibinfo {author} {\bibfnamefont {J.-X.}\ \bibnamefont {Lin}}, \bibinfo {author} {\bibfnamefont {Y.-H.}\ \bibnamefont {Zhang}}, \bibinfo {author} {\bibfnamefont {E.}~\bibnamefont {Morissette}}, \bibinfo {author} {\bibfnamefont {Z.}~\bibnamefont {Wang}}, \bibinfo {author} {\bibfnamefont {S.}~\bibnamefont {Liu}}, \bibinfo {author} {\bibfnamefont {D.}~\bibnamefont {Rhodes}}, \bibinfo {author} {\bibfnamefont {K.}~\bibnamefont {Watanabe}}, \bibinfo {author} {\bibfnamefont {T.}~\bibnamefont {Taniguchi}}, \bibinfo {author} {\bibfnamefont {J.}~\bibnamefont {Hone}},\ and\ \bibinfo {author} {\bibfnamefont {J.~I.~A.}\ \bibnamefont {Li}},\ }\bibfield  {title} {\bibinfo {title} {Spin-orbit–driven ferromagnetism at half moiré filling in magic-angle twisted bilayer graphene},\ }\href {https://doi.org/10.1126/science.abh2889} {\bibfield  {journal} {\bibinfo  {journal} {Science}\ }\textbf {\bibinfo {volume} {375}},\ \bibinfo {pages} {437–441} (\bibinfo {year} {2022})}\BibitemShut {NoStop}%
\bibitem [{\citenamefont {Trovatello}\ \emph {et~al.}(2022)\citenamefont {Trovatello}, \citenamefont {Piccinini}, \citenamefont {Forti}, \citenamefont {Fabbri}, \citenamefont {Rossi}, \citenamefont {De~Silvestri}, \citenamefont {Coletti}, \citenamefont {Cerullo},\ and\ \citenamefont {Dal~Conte}}]{Trovatello2022}%
  \BibitemOpen
  \bibfield  {author} {\bibinfo {author} {\bibfnamefont {C.}~\bibnamefont {Trovatello}}, \bibinfo {author} {\bibfnamefont {G.}~\bibnamefont {Piccinini}}, \bibinfo {author} {\bibfnamefont {S.}~\bibnamefont {Forti}}, \bibinfo {author} {\bibfnamefont {F.}~\bibnamefont {Fabbri}}, \bibinfo {author} {\bibfnamefont {A.}~\bibnamefont {Rossi}}, \bibinfo {author} {\bibfnamefont {S.}~\bibnamefont {De~Silvestri}}, \bibinfo {author} {\bibfnamefont {C.}~\bibnamefont {Coletti}}, \bibinfo {author} {\bibfnamefont {G.}~\bibnamefont {Cerullo}},\ and\ \bibinfo {author} {\bibfnamefont {S.}~\bibnamefont {Dal~Conte}},\ }\bibfield  {title} {\bibinfo {title} {Ultrafast hot carrier transfer in ws2/graphene large area heterostructures},\ }\bibfield  {journal} {\bibinfo  {journal} {npj 2D Materials and Applications}\ }\textbf {\bibinfo {volume} {6}},\ \href {https://doi.org/10.1038/s41699-022-00299-4} {10.1038/s41699-022-00299-4} (\bibinfo {year} {2022})\BibitemShut {NoStop}%
\bibitem [{\citenamefont {Stepanov}\ \emph {et~al.}(2021)\citenamefont {Stepanov}, \citenamefont {Xie}, \citenamefont {Taniguchi}, \citenamefont {Watanabe}, \citenamefont {Lu}, \citenamefont {MacDonald}, \citenamefont {Bernevig},\ and\ \citenamefont {Efetov}}]{Stepanov2021}%
  \BibitemOpen
  \bibfield  {author} {\bibinfo {author} {\bibfnamefont {P.}~\bibnamefont {Stepanov}}, \bibinfo {author} {\bibfnamefont {M.}~\bibnamefont {Xie}}, \bibinfo {author} {\bibfnamefont {T.}~\bibnamefont {Taniguchi}}, \bibinfo {author} {\bibfnamefont {K.}~\bibnamefont {Watanabe}}, \bibinfo {author} {\bibfnamefont {X.}~\bibnamefont {Lu}}, \bibinfo {author} {\bibfnamefont {A.~H.}\ \bibnamefont {MacDonald}}, \bibinfo {author} {\bibfnamefont {B.~A.}\ \bibnamefont {Bernevig}},\ and\ \bibinfo {author} {\bibfnamefont {D.~K.}\ \bibnamefont {Efetov}},\ }\bibfield  {title} {\bibinfo {title} {Competing zero-field chern insulators in superconducting twisted bilayer graphene},\ }\href {https://doi.org/10.1103/PhysRevLett.127.197701} {\bibfield  {journal} {\bibinfo  {journal} {Phys. Rev. Lett.}\ }\textbf {\bibinfo {volume} {127}},\ \bibinfo {pages} {197701} (\bibinfo {year} {2021})}\BibitemShut {NoStop}%
\bibitem [{\citenamefont {Tseng}\ \emph {et~al.}(2022)\citenamefont {Tseng}, \citenamefont {Ma}, \citenamefont {Liu}, \citenamefont {Watanabe}, \citenamefont {Taniguchi}, \citenamefont {Chu},\ and\ \citenamefont {Yankowitz}}]{Tseng2022}%
  \BibitemOpen
  \bibfield  {author} {\bibinfo {author} {\bibfnamefont {C.-C.}\ \bibnamefont {Tseng}}, \bibinfo {author} {\bibfnamefont {X.}~\bibnamefont {Ma}}, \bibinfo {author} {\bibfnamefont {Z.}~\bibnamefont {Liu}}, \bibinfo {author} {\bibfnamefont {K.}~\bibnamefont {Watanabe}}, \bibinfo {author} {\bibfnamefont {T.}~\bibnamefont {Taniguchi}}, \bibinfo {author} {\bibfnamefont {J.-H.}\ \bibnamefont {Chu}},\ and\ \bibinfo {author} {\bibfnamefont {M.}~\bibnamefont {Yankowitz}},\ }\bibfield  {title} {\bibinfo {title} {Anomalous hall effect at half filling in twisted bilayer graphene},\ }\href {https://doi.org/10.1038/s41567-022-01697-7} {\bibfield  {journal} {\bibinfo  {journal} {Nature Physics}\ }\textbf {\bibinfo {volume} {18}},\ \bibinfo {pages} {1038–1042} (\bibinfo {year} {2022})}\BibitemShut {NoStop}%
\bibitem [{\citenamefont {Bhowmik}\ \emph {et~al.}(2023)\citenamefont {Bhowmik}, \citenamefont {Ghawri}, \citenamefont {Park}, \citenamefont {Lee}, \citenamefont {Datta}, \citenamefont {Soni}, \citenamefont {Watanabe}, \citenamefont {Taniguchi}, \citenamefont {Ghosh}, \citenamefont {Jung},\ and\ \citenamefont {Chandni}}]{Bhowmik2023}%
  \BibitemOpen
  \bibfield  {author} {\bibinfo {author} {\bibfnamefont {S.}~\bibnamefont {Bhowmik}}, \bibinfo {author} {\bibfnamefont {B.}~\bibnamefont {Ghawri}}, \bibinfo {author} {\bibfnamefont {Y.}~\bibnamefont {Park}}, \bibinfo {author} {\bibfnamefont {D.}~\bibnamefont {Lee}}, \bibinfo {author} {\bibfnamefont {S.}~\bibnamefont {Datta}}, \bibinfo {author} {\bibfnamefont {R.}~\bibnamefont {Soni}}, \bibinfo {author} {\bibfnamefont {K.}~\bibnamefont {Watanabe}}, \bibinfo {author} {\bibfnamefont {T.}~\bibnamefont {Taniguchi}}, \bibinfo {author} {\bibfnamefont {A.}~\bibnamefont {Ghosh}}, \bibinfo {author} {\bibfnamefont {J.}~\bibnamefont {Jung}},\ and\ \bibinfo {author} {\bibfnamefont {U.}~\bibnamefont {Chandni}},\ }\bibfield  {title} {\bibinfo {title} {Spin-orbit coupling-enhanced valley ordering of malleable bands in twisted bilayer graphene on wse2},\ }\bibfield  {journal} {\bibinfo  {journal} {Nature Communications}\ }\textbf {\bibinfo {volume} {14}},\ \href {https://doi.org/10.1038/s41467-023-39855-x}
  {10.1038/s41467-023-39855-x} (\bibinfo {year} {2023})\BibitemShut {NoStop}%
\bibitem [{\citenamefont {Xie}\ \emph {et~al.}(2024)\citenamefont {Xie}, \citenamefont {Xu}, \citenamefont {Dong}, \citenamefont {Cui}, \citenamefont {Ou}, \citenamefont {Erdi}, \citenamefont {Watanabe}, \citenamefont {Taniguchi}, \citenamefont {Tongay}, \citenamefont {Levitov},\ and\ \citenamefont {Jin}}]{Xie2024}%
  \BibitemOpen
  \bibfield  {author} {\bibinfo {author} {\bibfnamefont {T.}~\bibnamefont {Xie}}, \bibinfo {author} {\bibfnamefont {S.}~\bibnamefont {Xu}}, \bibinfo {author} {\bibfnamefont {Z.}~\bibnamefont {Dong}}, \bibinfo {author} {\bibfnamefont {Z.}~\bibnamefont {Cui}}, \bibinfo {author} {\bibfnamefont {Y.}~\bibnamefont {Ou}}, \bibinfo {author} {\bibfnamefont {M.}~\bibnamefont {Erdi}}, \bibinfo {author} {\bibfnamefont {K.}~\bibnamefont {Watanabe}}, \bibinfo {author} {\bibfnamefont {T.}~\bibnamefont {Taniguchi}}, \bibinfo {author} {\bibfnamefont {S.~A.}\ \bibnamefont {Tongay}}, \bibinfo {author} {\bibfnamefont {L.~S.}\ \bibnamefont {Levitov}},\ and\ \bibinfo {author} {\bibfnamefont {C.}~\bibnamefont {Jin}},\ }\bibfield  {title} {\bibinfo {title} {Long-lived isospin excitations in magic-angle twisted bilayer graphene},\ }\href {https://doi.org/10.1038/s41586-024-07880-5} {\bibfield  {journal} {\bibinfo  {journal} {Nature}\ }\textbf {\bibinfo {volume} {633}},\ \bibinfo {pages} {77–82} (\bibinfo {year} {2024})}\BibitemShut
  {NoStop}%
\bibitem [{\citenamefont {Wagner}\ \emph {et~al.}(2022)\citenamefont {Wagner}, \citenamefont {Kwan}, \citenamefont {Bultinck}, \citenamefont {Simon},\ and\ \citenamefont {Parameswaran}}]{Wagner2022}%
  \BibitemOpen
  \bibfield  {author} {\bibinfo {author} {\bibfnamefont {G.}~\bibnamefont {Wagner}}, \bibinfo {author} {\bibfnamefont {Y.~H.}\ \bibnamefont {Kwan}}, \bibinfo {author} {\bibfnamefont {N.}~\bibnamefont {Bultinck}}, \bibinfo {author} {\bibfnamefont {S.~H.}\ \bibnamefont {Simon}},\ and\ \bibinfo {author} {\bibfnamefont {S.~A.}\ \bibnamefont {Parameswaran}},\ }\bibfield  {title} {\bibinfo {title} {Global phase diagram of the normal state of twisted bilayer graphene},\ }\href {https://doi.org/10.1103/PhysRevLett.128.156401} {\bibfield  {journal} {\bibinfo  {journal} {Phys. Rev. Lett.}\ }\textbf {\bibinfo {volume} {128}},\ \bibinfo {pages} {156401} (\bibinfo {year} {2022})}\BibitemShut {NoStop}%
\bibitem [{\citenamefont {Brei\o{}}\ and\ \citenamefont {Andersen}(2023)}]{Breio2023}%
  \BibitemOpen
  \bibfield  {author} {\bibinfo {author} {\bibfnamefont {C.~N.}\ \bibnamefont {Brei\o{}}}\ and\ \bibinfo {author} {\bibfnamefont {B.~M.}\ \bibnamefont {Andersen}},\ }\bibfield  {title} {\bibinfo {title} {Chern insulator phases and spontaneous spin and valley order in a moir\'e lattice model for magic-angle twisted bilayer graphene},\ }\href {https://doi.org/10.1103/PhysRevB.107.165114} {\bibfield  {journal} {\bibinfo  {journal} {Phys. Rev. B}\ }\textbf {\bibinfo {volume} {107}},\ \bibinfo {pages} {165114} (\bibinfo {year} {2023})}\BibitemShut {NoStop}%
\bibitem [{\citenamefont {Bultinck}\ \emph {et~al.}(2020)\citenamefont {Bultinck}, \citenamefont {Chatterjee},\ and\ \citenamefont {Zaletel}}]{Bultnick2020}%
  \BibitemOpen
  \bibfield  {author} {\bibinfo {author} {\bibfnamefont {N.}~\bibnamefont {Bultinck}}, \bibinfo {author} {\bibfnamefont {S.}~\bibnamefont {Chatterjee}},\ and\ \bibinfo {author} {\bibfnamefont {M.~P.}\ \bibnamefont {Zaletel}},\ }\bibfield  {title} {\bibinfo {title} {Mechanism for anomalous hall ferromagnetism in twisted bilayer graphene},\ }\href {https://doi.org/10.1103/PhysRevLett.124.166601} {\bibfield  {journal} {\bibinfo  {journal} {Phys. Rev. Lett.}\ }\textbf {\bibinfo {volume} {124}},\ \bibinfo {pages} {166601} (\bibinfo {year} {2020})}\BibitemShut {NoStop}%
\bibitem [{\citenamefont {Pitaevskii}(1960)}]{Pitaevskii1960}%
  \BibitemOpen
  \bibfield  {author} {\bibinfo {author} {\bibfnamefont {L.~P.}\ \bibnamefont {Pitaevskii}},\ }\bibfield  {title} {\bibinfo {title} {Electric forces in a transparent dispersive medium},\ }\href@noop {} {\bibfield  {journal} {\bibinfo  {journal} {Zh. Exp. Teor. Fiz.}\ }\textbf {\bibinfo {volume} {39}},\ \bibinfo {pages} {1450} (\bibinfo {year} {1960})},\ \bibinfo {note} {[Sov. Phys. JETP 12, 1008 (1961)]}\BibitemShut {NoStop}%
\bibitem [{\citenamefont {van~der Ziel}\ \emph {et~al.}(1965)\citenamefont {van~der Ziel}, \citenamefont {Pershan},\ and\ \citenamefont {Malmstrom}}]{Ziel1965}%
  \BibitemOpen
  \bibfield  {author} {\bibinfo {author} {\bibfnamefont {J.~P.}\ \bibnamefont {van~der Ziel}}, \bibinfo {author} {\bibfnamefont {P.~S.}\ \bibnamefont {Pershan}},\ and\ \bibinfo {author} {\bibfnamefont {L.~D.}\ \bibnamefont {Malmstrom}},\ }\bibfield  {title} {\bibinfo {title} {Optically-induced magnetization resulting from the inverse faraday effect},\ }\href {https://doi.org/10.1103/PhysRevLett.15.190} {\bibfield  {journal} {\bibinfo  {journal} {Phys. Rev. Lett.}\ }\textbf {\bibinfo {volume} {15}},\ \bibinfo {pages} {190} (\bibinfo {year} {1965})}\BibitemShut {NoStop}%
\bibitem [{\citenamefont {Pershan}\ \emph {et~al.}(1966)\citenamefont {Pershan}, \citenamefont {van~der Ziel},\ and\ \citenamefont {Malmstrom}}]{Pershan1966}%
  \BibitemOpen
  \bibfield  {author} {\bibinfo {author} {\bibfnamefont {P.~S.}\ \bibnamefont {Pershan}}, \bibinfo {author} {\bibfnamefont {J.~P.}\ \bibnamefont {van~der Ziel}},\ and\ \bibinfo {author} {\bibfnamefont {L.~D.}\ \bibnamefont {Malmstrom}},\ }\bibfield  {title} {\bibinfo {title} {Theoretical discussion of the inverse faraday effect, raman scattering, and related phenomena},\ }\href {https://doi.org/10.1103/PhysRev.143.574} {\bibfield  {journal} {\bibinfo  {journal} {Phys. Rev.}\ }\textbf {\bibinfo {volume} {143}},\ \bibinfo {pages} {574} (\bibinfo {year} {1966})}\BibitemShut {NoStop}%
\bibitem [{\citenamefont {Sharma}\ and\ \citenamefont {Balatsky}(2024)}]{Sharma2024}%
  \BibitemOpen
  \bibfield  {author} {\bibinfo {author} {\bibfnamefont {P.}~\bibnamefont {Sharma}}\ and\ \bibinfo {author} {\bibfnamefont {A.~V.}\ \bibnamefont {Balatsky}},\ }\bibfield  {title} {\bibinfo {title} {Light-induced orbital magnetism in metals via inverse faraday effect},\ }\href {https://doi.org/10.1103/PhysRevB.110.094302} {\bibfield  {journal} {\bibinfo  {journal} {Phys. Rev. B}\ }\textbf {\bibinfo {volume} {110}},\ \bibinfo {pages} {094302} (\bibinfo {year} {2024})}\BibitemShut {NoStop}%
\bibitem [{\citenamefont {Cheng}\ \emph {et~al.}(2020)\citenamefont {Cheng}, \citenamefont {Son},\ and\ \citenamefont {Sheldon}}]{Cheng2020}%
  \BibitemOpen
  \bibfield  {author} {\bibinfo {author} {\bibfnamefont {O.~H.-C.}\ \bibnamefont {Cheng}}, \bibinfo {author} {\bibfnamefont {D.~H.}\ \bibnamefont {Son}},\ and\ \bibinfo {author} {\bibfnamefont {M.}~\bibnamefont {Sheldon}},\ }\bibfield  {title} {\bibinfo {title} {Light-induced magnetism in plasmonic gold nanoparticles},\ }\href {https://doi.org/10.1038/s41566-020-0603-3} {\bibfield  {journal} {\bibinfo  {journal} {Nature Photonics}\ }\textbf {\bibinfo {volume} {14}},\ \bibinfo {pages} {365} (\bibinfo {year} {2020})}\BibitemShut {NoStop}%
\bibitem [{\citenamefont {Ortiz~Jimenez}\ \emph {et~al.}(2024)\citenamefont {Ortiz~Jimenez}, \citenamefont {Pham}, \citenamefont {Zhou}, \citenamefont {Liu}, \citenamefont {Nugera}, \citenamefont {Kalappattil}, \citenamefont {Eggers}, \citenamefont {Hoang}, \citenamefont {Duong}, \citenamefont {Terrones}, \citenamefont {Rodriguez~Gutiérrez},\ and\ \citenamefont {Phan}}]{Jimenez2024}%
  \BibitemOpen
  \bibfield  {author} {\bibinfo {author} {\bibfnamefont {V.}~\bibnamefont {Ortiz~Jimenez}}, \bibinfo {author} {\bibfnamefont {Y.~T.~H.}\ \bibnamefont {Pham}}, \bibinfo {author} {\bibfnamefont {D.}~\bibnamefont {Zhou}}, \bibinfo {author} {\bibfnamefont {M.}~\bibnamefont {Liu}}, \bibinfo {author} {\bibfnamefont {F.~A.}\ \bibnamefont {Nugera}}, \bibinfo {author} {\bibfnamefont {V.}~\bibnamefont {Kalappattil}}, \bibinfo {author} {\bibfnamefont {T.}~\bibnamefont {Eggers}}, \bibinfo {author} {\bibfnamefont {K.}~\bibnamefont {Hoang}}, \bibinfo {author} {\bibfnamefont {D.~L.}\ \bibnamefont {Duong}}, \bibinfo {author} {\bibfnamefont {M.}~\bibnamefont {Terrones}}, \bibinfo {author} {\bibfnamefont {H.}~\bibnamefont {Rodriguez~Gutiérrez}},\ and\ \bibinfo {author} {\bibfnamefont {M.-H.}\ \bibnamefont {Phan}},\ }\bibfield  {title} {\bibinfo {title} {Transition metal dichalcogenides: Making atomic-level magnetism tunable with light at room temperature},\ }\href {https://doi.org/https://doi.org/10.1002/advs.202304792}
  {\bibfield  {journal} {\bibinfo  {journal} {Advanced Science}\ }\textbf {\bibinfo {volume} {11}},\ \bibinfo {pages} {2304792} (\bibinfo {year} {2024})}\BibitemShut {NoStop}%
\bibitem [{\citenamefont {Argyres}(1955)}]{Argyres1955}%
  \BibitemOpen
  \bibfield  {author} {\bibinfo {author} {\bibfnamefont {P.~N.}\ \bibnamefont {Argyres}},\ }\bibfield  {title} {\bibinfo {title} {Theory of the faraday and kerr effects in ferromagnetics},\ }\href {https://doi.org/10.1103/PhysRev.97.334} {\bibfield  {journal} {\bibinfo  {journal} {Phys. Rev.}\ }\textbf {\bibinfo {volume} {97}},\ \bibinfo {pages} {334} (\bibinfo {year} {1955})}\BibitemShut {NoStop}%
\bibitem [{\citenamefont {Landau}\ \emph {et~al.}(1984)\citenamefont {Landau}, \citenamefont {Lifshitz},\ and\ \citenamefont {Pitaevskii}}]{LL1984}%
  \BibitemOpen
  \bibfield  {author} {\bibinfo {author} {\bibfnamefont {L.~D.}\ \bibnamefont {Landau}}, \bibinfo {author} {\bibfnamefont {E.~M.}\ \bibnamefont {Lifshitz}},\ and\ \bibinfo {author} {\bibfnamefont {L.~P.}\ \bibnamefont {Pitaevskii}},\ }\href@noop {} {\emph {\bibinfo {title} {{Electrodynamics of Continuous Media}}}},\ Vol.~\bibinfo {volume} {8}\ (\bibinfo  {publisher} {Pergamon Press, Oxford, England},\ \bibinfo {year} {1984})\ Chap.\ \bibinfo {chapter} {101}\BibitemShut {NoStop}%
\bibitem [{\citenamefont {Hertel}(2006)}]{Hertel2006}%
  \BibitemOpen
  \bibfield  {author} {\bibinfo {author} {\bibfnamefont {R.}~\bibnamefont {Hertel}},\ }\bibfield  {title} {\bibinfo {title} {Theory of the inverse faraday effect in metals},\ }\href {https://doi.org/https://doi.org/10.1016/j.jmmm.2005.10.225} {\bibfield  {journal} {\bibinfo  {journal} {Journal of Magnetism and Magnetic Materials}\ }\textbf {\bibinfo {volume} {303}},\ \bibinfo {pages} {L1} (\bibinfo {year} {2006})}\BibitemShut {NoStop}%
\bibitem [{\citenamefont {Rickhaus}\ \emph {et~al.}(2020)\citenamefont {Rickhaus}, \citenamefont {Liu}, \citenamefont {Kurpas}, \citenamefont {Kurzmann}, \citenamefont {Lee}, \citenamefont {Overweg}, \citenamefont {Eich}, \citenamefont {Pisoni}, \citenamefont {Taniguchi}, \citenamefont {Watanabe}, \citenamefont {Richter}, \citenamefont {Ensslin},\ and\ \citenamefont {Ihn}}]{Rickhaus2020}%
  \BibitemOpen
  \bibfield  {author} {\bibinfo {author} {\bibfnamefont {P.}~\bibnamefont {Rickhaus}}, \bibinfo {author} {\bibfnamefont {M.-H.}\ \bibnamefont {Liu}}, \bibinfo {author} {\bibfnamefont {M.}~\bibnamefont {Kurpas}}, \bibinfo {author} {\bibfnamefont {A.}~\bibnamefont {Kurzmann}}, \bibinfo {author} {\bibfnamefont {Y.}~\bibnamefont {Lee}}, \bibinfo {author} {\bibfnamefont {H.}~\bibnamefont {Overweg}}, \bibinfo {author} {\bibfnamefont {M.}~\bibnamefont {Eich}}, \bibinfo {author} {\bibfnamefont {R.}~\bibnamefont {Pisoni}}, \bibinfo {author} {\bibfnamefont {T.}~\bibnamefont {Taniguchi}}, \bibinfo {author} {\bibfnamefont {K.}~\bibnamefont {Watanabe}}, \bibinfo {author} {\bibfnamefont {K.}~\bibnamefont {Richter}}, \bibinfo {author} {\bibfnamefont {K.}~\bibnamefont {Ensslin}},\ and\ \bibinfo {author} {\bibfnamefont {T.}~\bibnamefont {Ihn}},\ }\bibfield  {title} {\bibinfo {title} {The electronic thickness of graphene},\ }\href {https://doi.org/10.1126/sciadv.aay8409} {\bibfield  {journal} {\bibinfo  {journal} {Science
  Advances}\ }\textbf {\bibinfo {volume} {6}},\ \bibinfo {pages} {eaay8409} (\bibinfo {year} {2020})}\BibitemShut {NoStop}%
\bibitem [{\citenamefont {Tschirhart}\ \emph {et~al.}(2021)\citenamefont {Tschirhart}, \citenamefont {Serlin}, \citenamefont {Polshyn}, \citenamefont {Shragai}, \citenamefont {Xia}, \citenamefont {Zhu}, \citenamefont {Zhang}, \citenamefont {Watanabe}, \citenamefont {Taniguchi}, \citenamefont {Huber},\ and\ \citenamefont {Young}}]{Tschirhart2021}%
  \BibitemOpen
  \bibfield  {author} {\bibinfo {author} {\bibfnamefont {C.~L.}\ \bibnamefont {Tschirhart}}, \bibinfo {author} {\bibfnamefont {M.}~\bibnamefont {Serlin}}, \bibinfo {author} {\bibfnamefont {H.}~\bibnamefont {Polshyn}}, \bibinfo {author} {\bibfnamefont {A.}~\bibnamefont {Shragai}}, \bibinfo {author} {\bibfnamefont {Z.}~\bibnamefont {Xia}}, \bibinfo {author} {\bibfnamefont {J.}~\bibnamefont {Zhu}}, \bibinfo {author} {\bibfnamefont {Y.}~\bibnamefont {Zhang}}, \bibinfo {author} {\bibfnamefont {K.}~\bibnamefont {Watanabe}}, \bibinfo {author} {\bibfnamefont {T.}~\bibnamefont {Taniguchi}}, \bibinfo {author} {\bibfnamefont {M.~E.}\ \bibnamefont {Huber}},\ and\ \bibinfo {author} {\bibfnamefont {A.~F.}\ \bibnamefont {Young}},\ }\bibfield  {title} {\bibinfo {title} {Imaging orbital ferromagnetism in a moiré chern insulator},\ }\href {https://doi.org/10.1126/science.abd3190} {\bibfield  {journal} {\bibinfo  {journal} {Science}\ }\textbf {\bibinfo {volume} {372}},\ \bibinfo {pages} {1323–1327} (\bibinfo {year}
  {2021})}\BibitemShut {NoStop}%
\bibitem [{\citenamefont {Grover}\ \emph {et~al.}(2022)\citenamefont {Grover}, \citenamefont {Bocarsly}, \citenamefont {Uri}, \citenamefont {Stepanov}, \citenamefont {Di~Battista}, \citenamefont {Roy}, \citenamefont {Xiao}, \citenamefont {Meltzer}, \citenamefont {Myasoedov}, \citenamefont {Pareek}, \citenamefont {Watanabe}, \citenamefont {Taniguchi}, \citenamefont {Yan}, \citenamefont {Stern}, \citenamefont {Berg}, \citenamefont {Efetov},\ and\ \citenamefont {Zeldov}}]{Grover2022}%
  \BibitemOpen
  \bibfield  {author} {\bibinfo {author} {\bibfnamefont {S.}~\bibnamefont {Grover}}, \bibinfo {author} {\bibfnamefont {M.}~\bibnamefont {Bocarsly}}, \bibinfo {author} {\bibfnamefont {A.}~\bibnamefont {Uri}}, \bibinfo {author} {\bibfnamefont {P.}~\bibnamefont {Stepanov}}, \bibinfo {author} {\bibfnamefont {G.}~\bibnamefont {Di~Battista}}, \bibinfo {author} {\bibfnamefont {I.}~\bibnamefont {Roy}}, \bibinfo {author} {\bibfnamefont {J.}~\bibnamefont {Xiao}}, \bibinfo {author} {\bibfnamefont {A.~Y.}\ \bibnamefont {Meltzer}}, \bibinfo {author} {\bibfnamefont {Y.}~\bibnamefont {Myasoedov}}, \bibinfo {author} {\bibfnamefont {K.}~\bibnamefont {Pareek}}, \bibinfo {author} {\bibfnamefont {K.}~\bibnamefont {Watanabe}}, \bibinfo {author} {\bibfnamefont {T.}~\bibnamefont {Taniguchi}}, \bibinfo {author} {\bibfnamefont {B.}~\bibnamefont {Yan}}, \bibinfo {author} {\bibfnamefont {A.}~\bibnamefont {Stern}}, \bibinfo {author} {\bibfnamefont {E.}~\bibnamefont {Berg}}, \bibinfo {author} {\bibfnamefont {D.~K.}\ \bibnamefont
  {Efetov}},\ and\ \bibinfo {author} {\bibfnamefont {E.}~\bibnamefont {Zeldov}},\ }\bibfield  {title} {\bibinfo {title} {Chern mosaic and berry-curvature magnetism in magic-angle graphene},\ }\href {https://doi.org/10.1038/s41567-022-01635-7} {\bibfield  {journal} {\bibinfo  {journal} {Nature Physics}\ }\textbf {\bibinfo {volume} {18}},\ \bibinfo {pages} {885–892} (\bibinfo {year} {2022})}\BibitemShut {NoStop}%
\bibitem [{\citenamefont {Stefani}\ \emph {et~al.}(2009)\citenamefont {Stefani}, \citenamefont {Hoogenboom},\ and\ \citenamefont {Barkai}}]{Stefani2009}%
  \BibitemOpen
  \bibfield  {author} {\bibinfo {author} {\bibfnamefont {F.~D.}\ \bibnamefont {Stefani}}, \bibinfo {author} {\bibfnamefont {J.~P.}\ \bibnamefont {Hoogenboom}},\ and\ \bibinfo {author} {\bibfnamefont {E.}~\bibnamefont {Barkai}},\ }\bibfield  {title} {\bibinfo {title} {Beyond quantum jumps: Blinking nanoscale light emitters},\ }\href {https://doi.org/10.1063/1.3086100} {\bibfield  {journal} {\bibinfo  {journal} {Physics Today}\ }\textbf {\bibinfo {volume} {62}},\ \bibinfo {pages} {34} (\bibinfo {year} {2009})}\BibitemShut {NoStop}%
\bibitem [{\citenamefont {Adhikari}\ \emph {et~al.}(2024)\citenamefont {Adhikari}, \citenamefont {Wang}, \citenamefont {Spaeth}, \citenamefont {Scalerandi}, \citenamefont {Albrecht}, \citenamefont {Liu},\ and\ \citenamefont {Orrit}}]{Adhikari2024}%
  \BibitemOpen
  \bibfield  {author} {\bibinfo {author} {\bibfnamefont {S.}~\bibnamefont {Adhikari}}, \bibinfo {author} {\bibfnamefont {Y.}~\bibnamefont {Wang}}, \bibinfo {author} {\bibfnamefont {P.}~\bibnamefont {Spaeth}}, \bibinfo {author} {\bibfnamefont {F.}~\bibnamefont {Scalerandi}}, \bibinfo {author} {\bibfnamefont {W.}~\bibnamefont {Albrecht}}, \bibinfo {author} {\bibfnamefont {J.}~\bibnamefont {Liu}},\ and\ \bibinfo {author} {\bibfnamefont {M.}~\bibnamefont {Orrit}},\ }\bibfield  {title} {\bibinfo {title} {Magnetization switching of single magnetite nanoparticles monitored optically},\ }\href {https://doi.org/10.1021/acs.nanolett.4c01850} {\bibfield  {journal} {\bibinfo  {journal} {Nano Letters}\ }\textbf {\bibinfo {volume} {24}},\ \bibinfo {pages} {9861} (\bibinfo {year} {2024})},\ \bibinfo {note} {pMID: 39078741}\BibitemShut {NoStop}%
\bibitem [{\citenamefont {Fisher}(1986)}]{Fisher1986}%
  \BibitemOpen
  \bibfield  {author} {\bibinfo {author} {\bibfnamefont {D.~S.}\ \bibnamefont {Fisher}},\ }\bibfield  {title} {\bibinfo {title} {Scaling and critical slowing down in random-field ising systems},\ }\href {https://doi.org/10.1103/PhysRevLett.56.416} {\bibfield  {journal} {\bibinfo  {journal} {Phys. Rev. Lett.}\ }\textbf {\bibinfo {volume} {56}},\ \bibinfo {pages} {416} (\bibinfo {year} {1986})}\BibitemShut {NoStop}%
\bibitem [{\citenamefont {Bittel}(1969)}]{Bittel1969}%
  \BibitemOpen
  \bibfield  {author} {\bibinfo {author} {\bibfnamefont {H.}~\bibnamefont {Bittel}},\ }\bibfield  {title} {\bibinfo {title} {Noise of ferromagnetic materials},\ }\href {https://doi.org/10.1109/tmag.1969.1066547} {\bibfield  {journal} {\bibinfo  {journal} {IEEE Transactions on Magnetics}\ }\textbf {\bibinfo {volume} {5}},\ \bibinfo {pages} {359–365} (\bibinfo {year} {1969})}\BibitemShut {NoStop}%
\bibitem [{\citenamefont {Carlson}\ \emph {et~al.}(2006)\citenamefont {Carlson}, \citenamefont {Dahmen}, \citenamefont {Fradkin},\ and\ \citenamefont {Kivelson}}]{Carlson2006}%
  \BibitemOpen
  \bibfield  {author} {\bibinfo {author} {\bibfnamefont {E.~W.}\ \bibnamefont {Carlson}}, \bibinfo {author} {\bibfnamefont {K.~A.}\ \bibnamefont {Dahmen}}, \bibinfo {author} {\bibfnamefont {E.}~\bibnamefont {Fradkin}},\ and\ \bibinfo {author} {\bibfnamefont {S.~A.}\ \bibnamefont {Kivelson}},\ }\bibfield  {title} {\bibinfo {title} {Hysteresis and noise from electronic nematicity in high-temperature superconductors},\ }\href {https://doi.org/10.1103/PhysRevLett.96.097003} {\bibfield  {journal} {\bibinfo  {journal} {Phys. Rev. Lett.}\ }\textbf {\bibinfo {volume} {96}},\ \bibinfo {pages} {097003} (\bibinfo {year} {2006})}\BibitemShut {NoStop}%
\bibitem [{\citenamefont {Bonetti}\ \emph {et~al.}(2004)\citenamefont {Bonetti}, \citenamefont {Caplan}, \citenamefont {Van~Harlingen},\ and\ \citenamefont {Weissman}}]{Bonetti2004}%
  \BibitemOpen
  \bibfield  {author} {\bibinfo {author} {\bibfnamefont {J.~A.}\ \bibnamefont {Bonetti}}, \bibinfo {author} {\bibfnamefont {D.~S.}\ \bibnamefont {Caplan}}, \bibinfo {author} {\bibfnamefont {D.~J.}\ \bibnamefont {Van~Harlingen}},\ and\ \bibinfo {author} {\bibfnamefont {M.~B.}\ \bibnamefont {Weissman}},\ }\bibfield  {title} {\bibinfo {title} {Electronic transport in underdoped ${\mathrm{y}\mathrm{b}\mathrm{a}}_{2}{\mathrm{c}\mathrm{u}}_{3}{\mathrm{o}}_{7\ensuremath{-}\ensuremath{\delta}}$ nanowires: Evidence for fluctuating domain structures},\ }\href {https://doi.org/10.1103/PhysRevLett.93.087002} {\bibfield  {journal} {\bibinfo  {journal} {Phys. Rev. Lett.}\ }\textbf {\bibinfo {volume} {93}},\ \bibinfo {pages} {087002} (\bibinfo {year} {2004})}\BibitemShut {NoStop}%
\bibitem [{\citenamefont {Deng}\ \emph {et~al.}(2020)\citenamefont {Deng}, \citenamefont {Ma}, \citenamefont {Wang}, \citenamefont {Yuan}, \citenamefont {Watanabe}, \citenamefont {Taniguchi}, \citenamefont {Zhang},\ and\ \citenamefont {Xia}}]{Deng2020}%
  \BibitemOpen
  \bibfield  {author} {\bibinfo {author} {\bibfnamefont {B.}~\bibnamefont {Deng}}, \bibinfo {author} {\bibfnamefont {C.}~\bibnamefont {Ma}}, \bibinfo {author} {\bibfnamefont {Q.}~\bibnamefont {Wang}}, \bibinfo {author} {\bibfnamefont {S.}~\bibnamefont {Yuan}}, \bibinfo {author} {\bibfnamefont {K.}~\bibnamefont {Watanabe}}, \bibinfo {author} {\bibfnamefont {T.}~\bibnamefont {Taniguchi}}, \bibinfo {author} {\bibfnamefont {F.}~\bibnamefont {Zhang}},\ and\ \bibinfo {author} {\bibfnamefont {F.}~\bibnamefont {Xia}},\ }\bibfield  {title} {\bibinfo {title} {Strong mid-infrared photoresponse in small-twist-angle bilayer graphene},\ }\href {https://doi.org/10.1038/s41566-020-0644-7} {\bibfield  {journal} {\bibinfo  {journal} {Nature Photonics}\ }\textbf {\bibinfo {volume} {14}},\ \bibinfo {pages} {549–553} (\bibinfo {year} {2020})}\BibitemShut {NoStop}%
\bibitem [{\citenamefont {Di~Battista}\ \emph {et~al.}(2024)\citenamefont {Di~Battista}, \citenamefont {Fong}, \citenamefont {Díez-Carlón}, \citenamefont {Watanabe}, \citenamefont {Taniguchi},\ and\ \citenamefont {Efetov}}]{DiBattista2024}%
  \BibitemOpen
  \bibfield  {author} {\bibinfo {author} {\bibfnamefont {G.}~\bibnamefont {Di~Battista}}, \bibinfo {author} {\bibfnamefont {K.~C.}\ \bibnamefont {Fong}}, \bibinfo {author} {\bibfnamefont {A.}~\bibnamefont {Díez-Carlón}}, \bibinfo {author} {\bibfnamefont {K.}~\bibnamefont {Watanabe}}, \bibinfo {author} {\bibfnamefont {T.}~\bibnamefont {Taniguchi}},\ and\ \bibinfo {author} {\bibfnamefont {D.~K.}\ \bibnamefont {Efetov}},\ }\bibfield  {title} {\bibinfo {title} {Infrared single-photon detection with superconducting magic-angle twisted bilayer graphene},\ }\bibfield  {journal} {\bibinfo  {journal} {Science Advances}\ }\textbf {\bibinfo {volume} {10}},\ \href {https://doi.org/10.1126/sciadv.adp3725} {10.1126/sciadv.adp3725} (\bibinfo {year} {2024})\BibitemShut {NoStop}%
\bibitem [{\citenamefont {Tokman}\ \emph {et~al.}(2020)\citenamefont {Tokman}, \citenamefont {Chen}, \citenamefont {Shereshevsky}, \citenamefont {Pozdnyakova}, \citenamefont {Oladyshkin}, \citenamefont {Tokman},\ and\ \citenamefont {Belyanin}}]{Tokman2020}%
  \BibitemOpen
  \bibfield  {author} {\bibinfo {author} {\bibfnamefont {I.~D.}\ \bibnamefont {Tokman}}, \bibinfo {author} {\bibfnamefont {Q.}~\bibnamefont {Chen}}, \bibinfo {author} {\bibfnamefont {I.~A.}\ \bibnamefont {Shereshevsky}}, \bibinfo {author} {\bibfnamefont {V.~I.}\ \bibnamefont {Pozdnyakova}}, \bibinfo {author} {\bibfnamefont {I.}~\bibnamefont {Oladyshkin}}, \bibinfo {author} {\bibfnamefont {M.}~\bibnamefont {Tokman}},\ and\ \bibinfo {author} {\bibfnamefont {A.}~\bibnamefont {Belyanin}},\ }\bibfield  {title} {\bibinfo {title} {Inverse faraday effect in graphene and weyl semimetals},\ }\href {https://doi.org/10.1103/PhysRevB.101.174429} {\bibfield  {journal} {\bibinfo  {journal} {Phys. Rev. B}\ }\textbf {\bibinfo {volume} {101}},\ \bibinfo {pages} {174429} (\bibinfo {year} {2020})}\BibitemShut {NoStop}%
\bibitem [{\citenamefont {Yoshino}(2011)}]{Yoshino2011}%
  \BibitemOpen
  \bibfield  {author} {\bibinfo {author} {\bibfnamefont {T.}~\bibnamefont {Yoshino}},\ }\bibfield  {title} {\bibinfo {title} {Simple theory of the inverse faraday effect with relationship to optical constants n and k},\ }\href {https://doi.org/https://doi.org/10.1016/j.jmmm.2011.05.010} {\bibfield  {journal} {\bibinfo  {journal} {Journal of Magnetism and Magnetic Materials}\ }\textbf {\bibinfo {volume} {323}},\ \bibinfo {pages} {2531} (\bibinfo {year} {2011})}\BibitemShut {NoStop}%
\bibitem [{\citenamefont {Tang}\ \emph {et~al.}(2019)\citenamefont {Tang}, \citenamefont {Mak},\ and\ \citenamefont {Shan}}]{Tang2019}%
  \BibitemOpen
  \bibfield  {author} {\bibinfo {author} {\bibfnamefont {Y.}~\bibnamefont {Tang}}, \bibinfo {author} {\bibfnamefont {K.~F.}\ \bibnamefont {Mak}},\ and\ \bibinfo {author} {\bibfnamefont {J.}~\bibnamefont {Shan}},\ }\bibfield  {title} {\bibinfo {title} {Long valley lifetime of dark excitons in single-layer wse2},\ }\href {https://doi.org/10.1038/s41467-019-12129-1} {\bibfield  {journal} {\bibinfo  {journal} {Nature Communications}\ }\textbf {\bibinfo {volume} {10}},\ \bibinfo {pages} {4047} (\bibinfo {year} {2019})}\BibitemShut {NoStop}%
\end{thebibliography}%

\newpage

\onecolumngrid
\newpage
\setcounter{section}{0}
\setcounter{figure}{0}
\renewcommand{\thefigure}{S\arabic{figure}}
\renewcommand{\theequation}{S.\arabic{equation}}
\renewcommand{\thetable}{S\arabic{table}}
\renewcommand{\thesection}{S\arabic{section}}

\renewcommand{\thefootnote}{\fnsymbol{footnote}}

\begin{center}

\textbf{SUPPLEMENTAL INFORMATION FOR\\
Optical control of orbital magnetism in magic angle twisted bilayer graphene}\\

\fontsize{9}{12}\selectfont

\vspace{1em}

Eylon Persky,$^{1,2,3*}$ Minhao He,$^4$ Jiaqi Cai,$^4$ Takashi Taniguchi,$^5$ Kenji Watanabe,$^6$ Xiaodong Xu,$^4$ and Aharon Kapitulnik$^{1,2,3,7}$

$^1${\it Geballe Laboratory for Advanced Materials, Stanford University, Stanford, California 94305, USA}\\
$^2${\it Stanford Institute for Materials and Energy Sciences, SLAC National Accelerator Laboratory, 2575 Sand Hill Road, Menlo Park, California 94025, USA}\\
$^3${\it Department of Applied Physics, Stanford University, Stanford, California 94305, USA}\\
$^4${\it Department of Physics, University of Washington, Seattle, Washington, 98195, USA}\\
$^5${\it Research Center for Materials Nanoarchitectonics,
National Institute for Materials Science, 1-1 Namiki, Tsukuba 305-0044, Japan.}\\
$^6${\it Research Center for Electronic and Optical Materials,
National Institute for Materials Science, 1-1 Namiki, Tsukuba 305-0044, Japan}\\
$^7${\it Department of Physics, Stanford University, Stanford, California 94305, USA}\\
\vspace{1em}

\end{center}

\section{Methods}
\noindent \textbf{Device fabrication.} The hetero-structure was assembled using a standard
dry-transfer technique with a PC/PDMS (polycarbonate/polydimethylsiloxane) stamp, and
transferred onto a Si/SiO$_2$ wafer. The twisted bilayer graphene was fabricated by using the tear-and-stack method. The top gate was a 3-5 layer thick graphene flake, in order to minimize its optical absorption. The back graphite gate was 3-5 nm thick. The Hall bar geometry was defined using CHF$_3$/O$_2$ and $O_2$ plasma etching, followed by electron beam lithography are used to define a Hall bar geometry. Cr/Au contacts (7nm/70nm) were added using electron beam evaporation. 
\bigskip

\noindent \textbf{Transport measurements.} The measurements were conducted in a Janis $^3$He cryostat. An a.c. current (1-5 nA rms, frequency 11 Hz) was applied to the sample, and the resulting voltage was read using a lock-in amplifier. A bottom gate voltage was applied using a Keithley 2450 in order to change the carrier density while the top gate was floating. 
\bigskip

\noindent \textbf{Optical setup.} Light was introduced into the cryostat via a 10 m polairzation maintaining fiber. A pair of aspheric lenses was used to collimate and refocus the light, resulting in a gaussian intensity profile incident on the sample, with a spot size that is determined by the ratio of the two focal lengths. In this study, a spot size of 4.5 $\mu$m was selected to match the dimensions of the device. A quarter wave plate was aligned at 45\textdegree\, with respect to the fast axis of the fiber, in order to convert light linearly polarized along the slow/fast axis to right/left circular polarization.

The lens tube was mounted on a piezo-based xy scanner to allow accurate positioning of the beam over the device at low temperatures. A piezo-based stick-slip positioner (Attocube ANPz101) was used to bring the sample into the focal plane of the lens. Finally, a manually-adjustable tip-tilt stage was used at room temperature to align the beam at normal incidence to the sample. At cryogenic temperatures, the lens tube was scanned over the device and the image of the reflected light intensity was compared to the optical microscope image of the device in order to then position the beam over the hetero-structure.
\bigskip

\section{Estimation of light-MATBG interaction}

In this section, we derive a relashipship between the Faraday effect and the inverse Faraday effect, which allows us to relate the magnetization induced by circularly polarized light to the rotation of linear polarization upon transmission through the magnetized material. We show that both effects can be expressed via the Verdet constant of the material. We use this relationship to show that the changes to $R_{xy}$ reported in Fig. \ref{fig-AHE_Pol} agree with the experimentally measured Faraday rotation.

Assuming an electric field $\textbf{E}(\omega)$ associated with circularly polarized light of total power $P(\omega)$ that is focused on an area $A$, which includes the sample. The intensity per unit area  that impinges on that area is given by the energy density of the electric field times the velocity of light
\begin{equation}
I_{\rm C}=\frac{P}{A}=\tfrac{1}{2}c\varepsilon_0|\textbf{E}(\omega)|^2
\label{exp}
\end{equation}
where $c$ is the speed of light,  $\varepsilon_0$ is the permittivity of the vacuum and the subscript C denotes the circular polarization right (C=R) or left (C=L). In our experiments about half the area of a beam with diameter 4.5 $\mu$m interacted with an active part of the sample (see Fig.~\ref{fig-photodoping}(b). Below we use two possible theoretical ideas to estimate the induced magnetic field due to circularly polarized light of total power $P=120 \ \mu$W focused on the sample.
\bigskip

\noindent \textbf{I. \ Inverse Faraday effect.}  The foundation to the theory of Inverse Faraday Effect was laid in the seminal paper of Pitaevskii \cite{Pitaevskii1960} on the response of a transparent dispersive medium to circularly polarized light and was further developed by Pershan {\it et al.} \cite{Ziel1965,Pershan1966}, who calculated the induced magnetization in zero magnetic field as a function of the material's so-called Verdet constant.  
\bigskip

\noindent \textbf{General Considerations:}

Starting from the energy density of the electromagnetic field in a dielectric medium without dispersion. In this case, the time-averaged energy density is:
\begin{equation}
u=\frac{1}{8\pi}\left[\textbf{E}\cdot\textbf{D}+\textbf{H}\cdot\textbf{B}\right]=\frac{1}{8\pi}\left[\varepsilon\textbf{E}^2+\mu\textbf{H}^2\right]
\end{equation}
From the first law of thermodynamics we know that:
\begin{equation}
du=Tds+\zeta d\rho-\textbf{H}\cdot d\textbf{B}/4\pi
\end{equation}
where $s$ is the entropy density, $\zeta$ is the chemical potential per unit mass and $\rho$ is the mass density (we use these instead of chemical potential and number density not to confuse with other parameters that carry similar notations in the discussion below). However, ultimately we will be interested in the case with no external magnetic field, but with possible finite magnetic induction. Therefore we use a Legendre transformation to define: $\tilde{u}=u-\textbf{H}\cdot\textbf{B}/4\pi$, such that
\begin{equation}
d\tilde{u}=Tds+\zeta d\rho-\textbf{B}\cdot d\textbf{H}/4\pi
\end{equation}
which implies that for a fixed density
\begin{equation}
\frac{\partial \tilde{u}}{\partial \textbf{H}}=-\frac{\textbf{B}}{4\pi}
\end{equation}
For the general case, assume an anisotropic material where the dielectric function in the material depends on magnetic field $\varepsilon_{ij}(\textbf{H})$. Let us denote the energy density of a medium without electric field as $u_0$. In the presence of an electric field the dielectric function may depend on the field. Let's assume now time varying fields. then we will be interested in time-averaged energy density, which will add a factor of 1/2 to the expression: 
\begin{equation}
-\frac{\textbf{B}}{4\pi}=\frac{\partial \tilde{u}}{\partial \textbf{H}}=\frac{\partial \tilde{u}_0}{\partial \textbf{H}}+\frac{\partial \varepsilon_{ij}}{\partial \textbf{H}}\frac{E_iE^*_j}{16\pi}
\label{eqnB}
\end{equation}
Taking $\mu=1$, we first note that:
\begin{equation}
\frac{\partial \tilde{u}_0}{\partial \textbf{H}}=-\frac{\textbf{H}}{4\pi}
\end{equation}
and with $\textbf{M}=(\textbf{B}-\textbf{H})/4\pi$ it implies:
\begin{equation}
-\textbf{M}=\frac{\partial \varepsilon_{ij}}{\partial \textbf{H}}\frac{E_iE^*_j}{16\pi}
\end{equation}

In the absence of magnetic field ($\textbf{H}=0$), and using complex time dependent electric field, which is circularly polarized: $\textbf{E}=E_0(\hat{x}+i\hat{y})e^{i(\vec{k}\cdot\vec{r}-\omega t)}$, we can show that
\begin{equation}
\textbf{M}=-\frac{\partial \varepsilon_{ij}}{\partial \textbf{H}}\Big|_{\textbf{H}=0}\frac{E_iE^*_j}{16\pi}\end{equation}
wich can be written in components as
\begin{equation}
M_l=-\frac{i}{16\pi}\epsilon_{ijk}\chi_{kl}E_iE_j^*
\end{equation}
where $\epsilon_{ijk}$ is the 3D totally antisymmetric tensor.  It is easy to check that relation for an isotropic medium (in particular if the circular polarization is in the $x$-$y$ plane and $\textbf{M}$ is in the $z$-direction), where $\chi_{kl}(\textbf{H}=0)=\chi \delta_{kl}$ and we can write:
\begin{equation}
\textbf{M}=-\chi\frac{1}{16\pi}(i\textbf{E}\times\textbf{E}^*)
\label{ifex}
\end{equation}

\noindent \textbf{The Faraday Effect:}

At the same time, let's assume an isotropic medium where the relation between the electric displacement- \textbf{D} and the electric field \textbf{E} in the presence magnetic field is:
\begin{equation}
\textbf{D}=\varepsilon\textbf{E}+i\textbf{E}\times\textbf{g}
\end{equation}
where $\textbf{g}=\chi\textbf{H}$. This in turn modify the index of refraction eigenvalues in the direction of propagation, which are now: $n^2_{\pm}=n_0^2\pm g$ where $\pm =$ R or L.  Let's look at the simple boundary problem of a linearly polarized wave moving in the $z$-direction in vacuum. We will represent the wave as a linear superposition of the two circular polarizations, thus the two respective wavevectors are $k_{\pm}=\omega n_{\pm}/c$. Obviously in vacuum  $\textbf{D}=\textbf{E}$ and $k_{\pm}=k=\omega/c$. Otherwise  we write:
\begin{equation}
D_x=\tfrac{1}{2} \left[ e^{i(k_{+} z)}+e^{i(k_{-}z)}\right], \ \ \ \ \  {\rm and} \ \ \ \ \ \  D_y=\tfrac{1}{2}\left[ -e^{i(k_{+}z)}+e^{i(k_{-}z)}\right],
\end{equation}
which can also be written as
\begin{equation}
D_x=\tfrac{1}{2}e^{ikz} \left[ e^{i\Delta k z}+e^{-i\Delta k z}\right]=e^{ikz} {\rm cos}(\Delta k z), \ \ \ \ \  {\rm and} \ \ \ \ \ \  D_y=\tfrac{1}{2}ie^{ikz} \left[ -e^{i\Delta k z}+e^{-i\Delta k z}\right]=e^{ikz} {\rm sin}(\Delta k z),
\end{equation}
where  $k=\tfrac{1}{2}(k_++k_-)$ and $\Delta k=\tfrac{1}{2}(k_+-k_-)$, which is the modification of the electric displacement vector due to the fact that in the material $k_+\neq k_-$. Upon leaving the material after propagating a distance d through its thickness, we find:
\begin{equation}
\frac{D_y}{D_x}={\rm tan} (\Delta k d) = {\rm tan}\left(\frac{d\omega g}{2cn_0}\right)\equiv {\rm tan}\theta_F
\end{equation}
or, define the Faraday angle:
\begin{equation}
\theta_F=\frac{d\omega g}{2cn_0}=VHd
\end{equation}
where we define the Verdet constant of the material as
\begin{equation}
V=\frac{\chi \omega}{2cn_0}
\end{equation}
Note that there is a factor of $4\pi$ difference between the definition of $\chi$ in Pershan's papers \cite{Pershan1966} and our derivation above, which originates in our definition of $\chi$ in Eqn.~\ref{ifex}. In our case:
\begin{equation}
\chi=\frac{2cn_0}{\omega}V \ \ \ \ \ \ \ \  {\rm and} \ \ \ \ \ \ \ \textbf{M}_{ind}=-\frac{cn_0}{8\pi\omega}V(i\textbf{E}\times\textbf{E}^*)
\end{equation}
Note that here we are interested in the material and therefore $\omega=c(2\pi/n_0\lambda_0)$, implying:
\begin{equation}
 \textbf{M}_{ind}=-\frac{n_0^2\lambda_0}{16\pi^2}V(i\textbf{E}\times\textbf{E}^*)
\end{equation}
We note that in our experiment we send light with power per unit area $P/A$ such that:
\begin{equation}
I_C=\frac{P}{A}=\frac{1}{2}c(n_0^2\varepsilon_0)|E|^2=\frac{1}{2}c\frac{n_0^2}{4\pi}|E|^2
\end{equation}
where in the second equality we use the cgs substitution: $\varepsilon_0=1/4\pi$. The result is then \cite{Pershan1966}:
\begin{equation}
M_{ind}=\frac{\lambda_0}{2\pi c}VI_C=\frac{\lambda_0}{2\pi c}\frac{\theta_F}{Hd}I_C
\label{ind}
\end{equation}
\bigskip

\noindent \textbf{Comparison to Experiment:}

In the above derivation we rely on a measurement of the Faraday angle, which can allow us to calculate the projected induced magnetization. In our experiments on MATBG we measured a ``double faraday effect'' due to the fact that the sample is very thin an the substrate behind serves as a mirror. To extract an order of magnitude of the expected induced magnetization, we may assume that $\theta_F\approx 10 \ \mu$rad at a field of 100 Oe (see Fig.~\ref{fig_KF}~below). We are interested in induced magnetization for light of intensity 120 $\mu$W at a spot size diameter of $\sim4.5 \ \mu$m. For a light source with $\lambda_0=1.55 \ \mu$m we obtain $M_{ind}\approx 1\times 10^{-5}$ erg/Oe$\cdot$cm$^3$. It is instructive to see how many Bohr magneton it implies, so we divide by $\mu_B\approx 10^{-20}$ erg/Oe, which yields: $M_{ind}\approx 1\times 10^{15}\mu_B/{\rm cm}^3$. Multiplying by the thickness (we take $d\approx 5\times10^{-8}$ cm for the bilayer graphene, we obtain for the area induced moment: $M^{2D}_{ind}\approx 5\times 10^7 \ \mu_B/{\rm cm}^2$. A moir\'e unit cell has an area of $\sim(10 {\rm nm})^2 \sim 10^{-12} \ {\rm cm}^2$, which implies $\sim 5\times 10^{-5}\mu_B$ per moir\'e unit cell - a very small moment.\begin{figure*}[ht!]
\includegraphics[width=0.5\columnwidth]{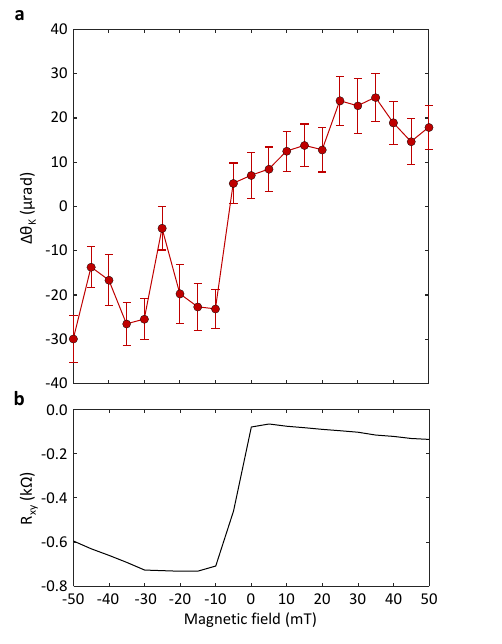}
\caption{Kerr/Faraday Effect hysteresis measurement of MATBG. Note: In our measurement if the substrate serves as a mirror and Faraday effect dominates, the TRSB signal is $4\times\theta_F$. The error bars represent 95\% confidence intervals obtained from averaging over time.}
\label{fig_KF}
\end{figure*}
\bigskip

Note that if for example the moments came out of electrons excited from the flat band and found a remote band with long lifetime, then we know that our level of photo-doping is about 2\% or less. Since at $\nu=1$ the carrier density in the flat band is approx. $1x10^{12}$ per cm$^2$, taking 2\%, we find that we may induce about $2\times10^{10} \ \mu_B$/cm$^2$., about two orders of magnitude more than in the IFE scenario. However, this must be a vast overestimate since we need the difference in photo-doping between light of right and left circularly polarized light, which may be much smaller.
\bigskip

\noindent \textbf{Semi-classical estimation of the Verdet constant}

The Inverse faraday effect can also be calculated in a semi-classical approach from simple transport equations. This is done by noting that the circularly polarized light induces a circular electric field, which the electrons need to follow. Following Hertel  \cite{Hertel2006}  we find that  the current density that induces solenoidal magnetization is:
\begin{equation}
\textbf{j}_s=\frac{i}{4e\langle n\rangle \omega}\nabla\times\left[(\sigma\textbf{E})\times(\sigma\textbf{E})^*\right],
\end{equation}
which induces a magnetic moment through $\textbf{j}_s=\nabla \times \textbf{M}_{ind}$ of magnitude:
\begin{equation}
\textbf{M}_{ind}=\frac{1}{e\langle n\rangle \varepsilon_0\omega}\left[i\varepsilon_0(\sigma\textbf{E})\times(\sigma\textbf{E})^*\right],
\end{equation}
For itinerant electrons with no dissipation $\sigma=i\langle n\rangle e^2/m^*\omega$ to yield:
\begin{equation}
\textbf{M}_{ind}=\frac{e^3\langle n\rangle}{4m_*^2 \varepsilon_0\omega^3}\left[i\varepsilon_0\textbf{E}\times\textbf{E}^*\right]
\end{equation}
If we use the expression for density vs. Fermi velocity for 2D Dirac electrons in graphene, together with an effective thickness $d_g$ and 2-spin and 2-valley degeneracy total of 4: $d_g\langle n\rangle=2m_*^2v_F^2/\pi\hbar^2$, we obtain the known result in  \cite{Tokman2020}:
\begin{equation}
M_{\rm ind}=\frac{e^3 v_F^2}{2\pi \varepsilon_0 \hbar^2 \omega^3 d_g}(i\varepsilon_0\textbf{E}\times\textbf{E}^*)
\label{eqn-ife}
\end{equation}

Revisiting the IFE in the presence of a resonance (see e.g \cite{Yoshino2011}), The current associated with the circularly polarized electric field will be modified through  the optical conductivity of the resonance, which is described by a resonance frequency $\omega_0$ and lifetime $\tau=1/\Gamma$. The result for the induced magnetization is now modified by the factor $1/\omega^3 \to \omega/[(\omega^2-\omega_0^2)^2+\Gamma^2\omega^2]$ to yield:
\begin{equation}
\textbf{M}_{ind}=\frac{e^3v_F^2}{2 \pi\hbar^2\varepsilon_0d_g}\frac{\omega}{(\omega^2-\omega_0^2)^2+\Gamma^2\omega^2}\left[i\varepsilon_0\textbf{E}\times\textbf{E}^*\right]
\end{equation}
where if we set $\omega_0=\Gamma=0$ we are back at equation~\ref{eqn-ife}.

The case of WSe$_2$ over MATBG is interesting because due to inversion symmetry breaking our broad-band source of $1550\pm 30$ nm wavelength (0.785 to 0.816 eV) light gives a substantial second harmonic generation intensity which can resonate with the WSe$_2$ excitons with energies in the range 1.6 to 1.7 eV \cite{Tang2019,He2024}. In fact, Tang {\it et al.} showed a substantial excitonic yield in the energy range of 1.6 eV and 1.63 eV, which therefore has an overlap with our broadband SLED.  Assume that within our broadband illumination a fraction of photons $f$ are at resonance with the WSe$_2$ excitons, which have been shown to have a long lifetime, of order 10 ns, for either circular polarization \cite{Tang2019}. In that case the induced moment will be dominanted by
\begin{equation}
\textbf{M}_{ind}=f \frac{e^3v_F^2}{2 \pi\hbar^2\varepsilon_0d_g\omega\Gamma^2}\left[i\varepsilon_0\textbf{E}\times\textbf{E}^*\right]
\end{equation}
Taking $\Gamma\sim10^8$ s$^{-1}$ and $f\sim 0.5\%$, which is the fraction of the normal distribution of our SLED that can resonate with the excitons, we obtain a multiplication factor of $\sim 10^{10}$, which can now explain the observed IFE quantitatively.
\bigskip

\section{Additional Data}
\begin{figure}[h]
\includegraphics[width=1.0\columnwidth]{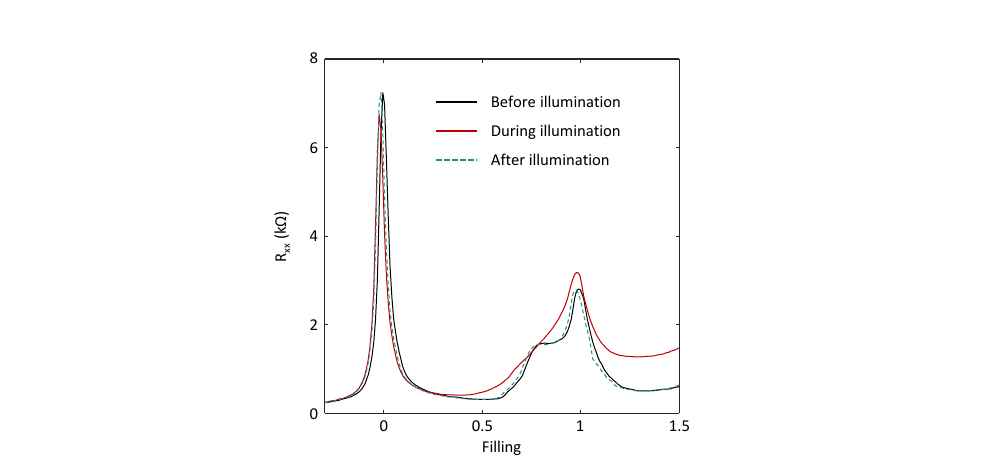}
\caption{Measurements of $R_{xx}$ before and after illuminating the sample, showing that the effect of the light is not persistent.}
\label{fig-nomemoryRxx}
\end{figure}

\begin{figure}[h]
\includegraphics[width=1.0\columnwidth]{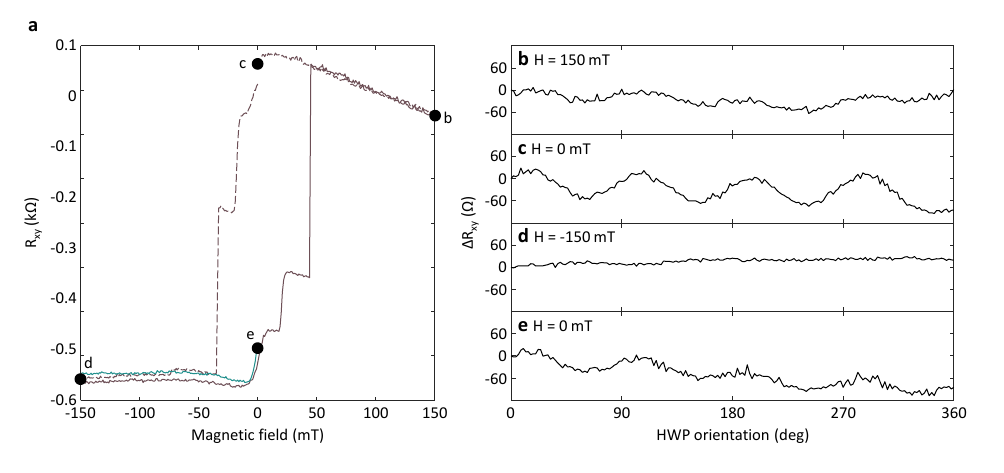}
\caption{(a) A hysteresis loop taken at $\nu = 0.95$ under illumination. (b) $\Delta R_{xy}$ as a function of the half waveplate orientation, taken at different points along the hysteresis loop in a. The oscillations were smaller at fields where $R_{xy}$ was saturated.}
\label{fig-S_HWP_hyst}
\end{figure}

\begin{figure}[h]
\includegraphics[width=1.0\columnwidth]{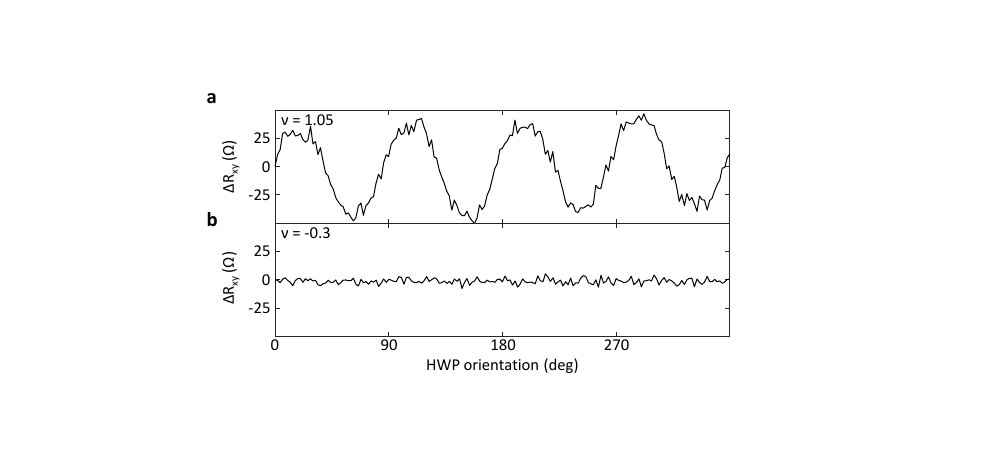}
\caption{$\Delta R_{xy}$ as a function of the half waveplate orientation near (a) and far away (b) from $\nu=1$.}
\label{fig-no_osc_03}
\end{figure}

\end{document}